\documentclass{article}
\usepackage{graphicx} % Required for inserting images
\usepackage{url} %this package should fix any errors with URLs in refs.
\usepackage{subfiles}
\usepackage{lineno}
\usepackage{amsmath}
\usepackage{amssymb}
\usepackage{mathrsfs}
\usepackage{amsmath}
\usepackage{appendix}
\usepackage{graphicx} % Added by Shoubhik Mondal, 2014.
\usepackage{biblatex}
\usepackage{dsfont}
\usepackage{float}
\usepackage{diagbox} 
\usepackage{subcaption}

\usepackage{graphics}
\usepackage{csquotes}
\usepackage[all]{xy}
\PassOptionsToPackage{hyphens}{url}
\usepackage{tikz-cd}
\usepackage{braket}
\usepackage{xcolor}
\usepackage{verbatim}
\usepackage{float}
\usepackage{mathrsfs}
\usepackage{extarrows}
\usepackage{tikz}
\usepackage[textwidth=15cm,textheight= 22cm]{geometry}
\usetikzlibrary{shapes,arrows,positioning}
\usepackage{mathrsfs}
\usepackage{mathtools}
\usepackage{hyperref}

\usepackage{wrapfig}
\usepackage{algorithm} 
\usepackage{algpseudocode} 
\usepackage{fancyhdr}
\usepackage{soul}

\newcommand{\bk}{\mathbf k}

\newcommand{\bu}{\mathbf u}

\newtheorem{rem}{Remark}[section]

\title{Accelerating Simulations of Tropical Cyclones using Adaptive Mesh Refinement}

\author{Yassine Tissaoui\footnote{Department of Mechanical and Industrial Engineering, New Jersey Institute of Technology, Newark NJ, U.S.A.. yt277@njit.edu}, Stephen R. Guimond\footnote{Department of Atmospheric and Planetary Sciences, Hampton University, Hampton VA, U.S.A.. stephen.guimond@hamptonu.edu}, Francis X. Giraldo\footnote{Department of Applied Mathematics, Naval Postgraduate School, Monterey, CA, U.S.A.. fxgirald@nps.edu}, Simone Marras\footnote{Department of Mechanical and Industrial Engineering, and Center for Applied Mathematics and Statistics, New Jersey Institute of Technology, Newark NJ, U.S.A.. smarras@njit.edu}}
\date{October 2024}

\begin{document}

\maketitle

\begin{abstract} 
Tropical cyclones (TCs) are powerful, natural phenomena that can severely impact populations and infrastructure. Enhancing our understanding of the mechanisms driving their intensification is crucial for mitigating these impacts. To this end, researchers are pushing the boundaries of TC simulation resolution down to scales of just a few meters. However, higher resolution simulations come with significant computational challenges, increasing both time and energy costs. Adaptive mesh refinement (AMR) is a technique widely used in computational fluid dynamics but has seen limited application in atmospheric simulations. This study explores the use of h-adaptive grids using the spectral element discretization technique to accelerate TC simulations while allowing very high resolutions in certain parts of the domain. By applying AMR to a rapidly intensifying TC test case, we demonstrate that AMR can replicate the results of uniform grid simulations in terms of mean and local wind speed maxima while dramatically reducing computational costs. We show that AMR can speed up dry TC simulations by a factor of 2 - 13 for the set of tested refinement criteria. Additionally, we show that TC intensity changes as resolution is increased and that AMR can deliver high-resolution simulations at the cost of coarser static simulations. Our findings indicate that AMR and spectral element methods are promising tools for enhancing TC simulations.
\end{abstract}

%%%%%%%%%%%%%%%%%%%%%%%%%%%%%%%%%%%%%%%%%%%%%%%%%%%%%%%%%%%%%%%%%%%%%
% SIGNIFICANCE STATEMENT/CAPSULE SUMMARY
%%%%%%%%%%%%%%%%%%%%%%%%%%%%%%%%%%%%%%%%%%%%%%%%%%%%%%%%%%%%%%%%%%%%%
%
% If you are including an optional significance statement for a journal article or a required capsule summary for BAMS 
% (see www.ametsoc.org/ams/index.cfm/publications/authors/journal-and-bams-authors/formatting-and-manuscript-components for details), 
% please apply the necessary command as shown below:

%\capsule
%       Enter BAMS capsule here, no more than 30 words. See \url{www.ametsoc.org/index.cfm/ams/publications/author-information/formatting-and-manuscript-components/#capsule} for details.
%
%% * * If using twocol mode, you will need to use the commands "twocolsig" and "twocolcapsule" in place of "sig" and "capsule"
%%      to ensure that the text box correctly spans across both columns.
%

%%%%%%%%%%%%%%%%%%%%%%%%%%%%%%%%%%%%%%%%%%%%%%%%%%%%%%%%%%%%%%%%%%%%%
% MAIN BODY OF PAPER
%%%%%%%%%%%%%%%%%%%%%%%%%%%%%%%%%%%%%%%%%%%%%%%%%%%%%%%%%%%%%%%%%%%%%
%
\section{Introduction}

Atmospheric phenomena can be challenging to simulate with mathematical models given the large range of scales that are present, which span several orders of magnitude, and the strong nonlinear interactions between those scales \cite{klein_2010, klein2010scale}. Tropical cyclones (TCs) are at the top of the list of challenging phenomena, given their extremely large Reynolds number ($\sim$ $10^{10}$) which requires large domains with high resolution and high order discretization to accurately capture the vast array of interactions controlling the system intensity. Physical processes such as (i) vortex Rossby waves \cite{Guimond2010,Guinn, WANG2002,Montgomery1997}, (ii) mesovortices at the eye/eyewall interface \cite{Schubert1999,Kossin2001, Hendricks2009,Guimond2016a, Cram2007}, (iii) boundary-layer coherent turbulent structures \cite{Guimond2018b, Foster2005} , (iv) air-sea turbulent fluxes \cite{WU2005,Emanuel1986}, and (v) deep convective bursts \cite{Rogers2013,Molinari2010,Hendricks2004,Guimond2010} all play a role in TC intensification and require high fidelity numerical treatment to accurately capture their energetics \cite{Badrul2022}. Increasing the model resolution and numerical accuracy are the primary means to reduce both the model error from parameterization and the error from excessive diffusion \cite{johnson2024impacts}. The computational burden of increasing the model resolution and numerical accuracy for explicitly simulating large turbulent eddies in TCs is substantial, even with the large core counts and optimized hardware of modern computing systems.  

Scientists and engineers have developed various techniques to alleviate the computational burden of high resolution simulations of atmospheric phenomena including:  (1) grid nesting and low-order discretizations (e.g., \cite{Kuriharaetal79,Kuriharaetal98,hodur1997,wrfModelDescription2007,Doyleetal2014}); (2) static grid stretching (e.g., \cite{Guimondetal2012}); and (3) adaptive mesh refinement or AMR (e.g., \cite{skamarock1989, kopera2014mass, abdi2024comparison}). Each of these techniques has positive and negative attributes. For example, static grid stretching is a relatively simple method to implement in a code, but it doesn't always reduce the computational cost effectively.  Grid nesting has been effective for zooming into a smaller region with higher resolution while leaving surrounding areas with coarser resolution, but noise at the nest boundaries and some difficulty following features of interest are apparent \cite{Fergusonetal2016}. The AMR method works by dynamically adjusting the grid resolution in a series of refinements according to a set of pre-defined criteria such as the wind speed or vorticity. In this case, the grid becomes adaptive by refining and coarsening as targeted flow features appear and vanish over time. 

The early work of \cite{bergerOliger1984}, \cite{skamarockEtAl1989}, and \cite{skamarockKlemp1993} showed that adaptive refinement is useful for solving hyperbolic equations and problems in numerical weather prediction. \cite{OMEGAmodel} developed the first operational model making use of AMR to generate horizontally adaptive grids. This model was successful in simulating hurricane storm tracks. AMR has been shown to be effective for simulations of the shallow water equations \cite{Bergeretal2009,leveque2011tsunami,McCorquodaleetal2015} and it has been tested with high-order Galerkin methods for atmospheric applications \cite{mueller_2013,kopera2014analysis,Chenetal2011}. Recent applications of AMR led to the improvement of modeling atmospheric flows around topography \cite{orlando2024robust,yamazakiEtAl2022,liEtAl2021}. 

In terms of TC simulations, \cite{hendricksEtAl2016} demonstrated that AMR can speed up highly idealized simulations of TCs by a factor of $4-15$ using a shallow water spectral element model. The goal of the current paper is to extend these results to realistic, three-dimensional TC simulations using a similar spectral element model that provides high-order accurate discretizations of the governing equations. The authors will quantify the computational advantage of AMR over uniform resolution simulations. These numerical methods are applied to the problem of TC rapid intensification, which is characterized by the pulsing of deep convective towers for $\sim$ 6 - 12 h time periods. The thermal forcing associated with these convective towers generates a spectrum of smaller-scale motions that serve as an excellent test case for the AMR algorithm.

%{\color{red} Add this citation somewhere: \cite{johnson2024impacts}}
 
\section{Model Equations}

We consider a dry atmosphere of density $\rho$, pressure $p$ and potential temperature $\theta = T/\pi$, where $T$ is the sensible temperature and $\pi$ is the exner pressure. Taking the ideal gas constant of dry air to be $R_d$ and its specific heat capacity at constant pressure to be $c_p$, the exner pressure is defined as $\pi = \left(p/p_0\right)^{R_d/c_p}$, where $p_0$ is the sea surface pressure. Let $\bf{u}$ be the wind velocity vector. We consider a fixed spatial domain $\Omega$ and a time interval of interest $(0, tf]$. We write the compressible Euler equations in non-conservative form as follows:

\begin{align}
    &\frac{\partial \rho}{\partial t} + \nabla \cdot (\rho \bu)= 0 &&\text{in } \Omega \times (0,t_f], \label{eq:mass}  \\
    & \frac{\partial \bu}{\partial t} + \bu \cdot \nabla \bu = - \frac{1}{\rho} \nabla p  + g\bk  &&\text{in } \Omega \times (0,t_f],  \label{eq:mom} \\
& \frac{\partial \theta}{\partial t} + \bu \cdot \nabla \theta = 0 &&\text{in } \Omega \times (0,t_f],
\label{eq:ent}
\end{align}
where $g=9.81$ ${\rm m/s^2}$ is the magnitude of the acceleration due to gravity and $\bk=[0, 0, -1]$ is the unit vector aligned with the vertical axis. The ideal gas law applied to dry air is used to close the system, 

\begin{equation}
    p = \rho R_d T.
\end{equation}

To facilitate the numerical solution of system \eqref{eq:mass}-\eqref{eq:ent}, we write density, pressure, 
and potential temperature as the sum of
their mean hydrostatic values and fluctuations:
\begin{align}
\rho(x,y,z,t) &= \rho_0(z) + \rho'(x,y,z,t), \label{eq:rho_split} \\
\theta(x,y,z,t) &= \theta_0(z) + \theta'(x,y,z,t), \label{eq:theta_split} \\
p(x,y,z,t) &= p_0(z) + p'(x,y,z,t). \label{eq:p_split}
\end{align}
Note that the hydrostatic reference states are functions
of the vertical coordinate $z$ only. Hydrostatic balance relates
$p_0$ to $\rho_0$ as follows:
\begin{align}
\frac{dp_0}{dz} = - \rho_0 g. \label{eq:hydro_bal}
\end{align}

Plugging \eqref{eq:rho_split}-\eqref{eq:p_split} into 
\eqref{eq:mass}-\eqref{eq:ent} and accounting for \eqref{eq:hydro_bal} leads to:

\begin{align}
& \frac{\partial \rho'}{\partial t}  + \nabla \cdot ((\rho_0+ \rho') \bu) = 0 &&\text{in } \Omega \times (0,t_f], \label{eq:mass_split} \\
& \frac{\partial \bu}{\partial t} + \bu \cdot \nabla \bu = - \frac{1}{\rho_0 + \rho'} \nabla p'  +\frac{\rho'}{\rho_0 + \rho'}g\bk &&\text{in } \Omega \times (0,t_f],\label{eq:mom_split} \\
& \frac{\partial \theta'}{\partial t} + \bu \cdot \nabla \theta_0 + \bu \cdot \nabla \theta' =0 &&\text{in } \Omega \times (0,t_f].
\label{eq:ent_split}
\end{align}

\section{The Spectral Element Method}

The continuous Galerkin (CG) spectral element method is used to discretize the governing equations in space. The three-dimensional (3D) basis functions for these elements are constructed using tensor products of the one-dimensional Lagrange interpolating polynomials $h$ of order $N$:
\begin{equation}
    \psi_l(\mathbf{x}) = h_i(\xi(\mathbf{x})) \otimes h_j(\eta(\mathbf{x})) \otimes h_k(\zeta(\mathbf{x})), \quad l=i +(N+1)(j + k(N+1)),
\end{equation}
where $\mathbf{x}=(x,y,z)$ and $\xi$, $\eta$ and $\zeta$ are mappings from the physical coordinate $\mathbf{x}$ onto each coordinate of the reference element. The Legendre Gauss Lobatto points are used as both the interpolation and integration points. We use spectral elements of order 4. This is a sufficiently high order for inexact integration to be accurate and allow for important computational cost savings. For details on the CG  method we refer the interested reader to \cite{KoprivaBook,giraldoBOOK}.

\subsection{Considerations for adaptive mesh refinement}
\begin{figure}
    \centering
    \includegraphics[width=\textwidth]{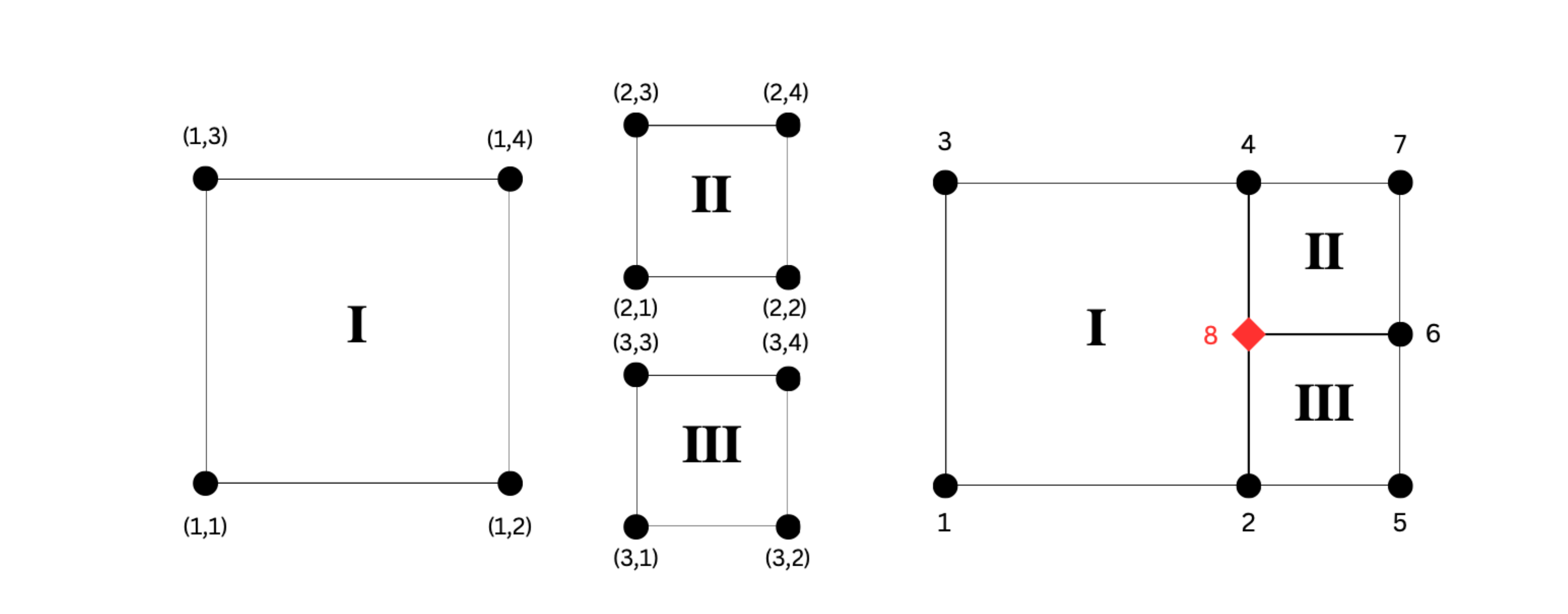}
    \caption{Illustration of non-conforming linear elements in local view (left) and global view (right). While node number 8 is shared by elements \textbf{II} and \textbf{III} corresponding to local nodes $(2,1)$ and $(3,3)$, respectively, it does not have a corresponding local node on element \textbf{I}. Node 8 is thus a hanging node.}
    \label{fig:hanging nodes}
\end{figure}

The use of adaptive mesh refinement (AMR) leads to the presence of non-conforming elements. That is to say that it is possible to obtain neighboring elements that do not necessarily have matching nodes along the face or edge that they share. We use linear elements to illustrate this in Figure \ref{fig:hanging nodes} where node 8 in the global view (right) corresponds to the local nodes $(2,1)$ and $(3,3)$ of elements \textbf{II} and \textbf{III}, respectively. However, this node has no corresponding local node on element \textbf{I} and is called a hanging node. 

The global solution of the CG method is constructed via direct stiffness summation (DSS) which constructs the solution at each global node based on the values of the solution at the corresponding local nodes of each element that shares the global node.  A modified DSS approach must be used to appropriately construct the global solution when there are hanging nodes. Essentially the values of the local solution at hanging nodes contribute to the global solution through interpolation onto non-hanging nodes. The mass matrix is also constructed through this approach which insures the correctness of the DSS operation. Once a global solution is obtained it is then interpolated onto the hanging nodes to construct the correct local solutions. We refer the reader to Section 2.4 of \cite{kopera2014mass} for algorithmic details.

\section{Adaptive Mesh Refinement (AMR)}

We rely on the \textbf{P4est} \cite{BursteddeWilcoxGhattas11} library to handle domain decomposition and load balancing of our parallel applications. This library also allows for refining and coarsening grids and we make use of this capability to perform adaptive mesh refinement during our simulations. 

\subsection{The AMR Procedure}
Thanks to \textbf{P4est} the AMR procedure is straightforward to perform. The process only requires that \textbf{P4est} know if an element should be refined, coarsened, or left untouched. For this to take place, an array ${\tt adapt}$ of size $Ne$, the total number of elements,  is sent to \textbf{P4est}. The values of this array for each element $e$ are as follows:
\begin{equation}
    {\tt adapt}(e) = \begin{cases}
        -1 & \textrm{If the element is to be marked for coarsening} \\
        0 & \textrm{If the element is to be remain as is} \\
        1  & \textrm{If the element is to be marked for refinement}
    \end{cases}
\end{equation}

\paragraph{Refining an element:}
In this work, we are only considering adaptive mesh refinement in the horizontal directions. As such, if a hexahedral element is marked for refinement, it is split into four elements of equal size in the horizontal plane. Since we depend on column data-structures for the planetary boundary layer package being used in our simulations, if an element belonging to a column is refined, the entire column must be refined with it to maintain this structure. 

\paragraph{Coarsening an element:}
If the refinement process involves splitting an element into smaller elements then the coarsening process is the opposite. Four neighboring elements that are marked for coarsening are merged together to form a single larger element. As such, coarsening cannot take place unless there are four adjacent elements that all share a corner.

\paragraph{Conditions for refinement and coarsening:} 
$\textbf{P4est}$ also stores the current level of refinement of each element in an array we will call ${\tt lvl}$. Initially, all elements have the level $\tt lvl(e) = 0$ indicating no refinement has taken place. This is also the maximum size of a given element, meaning that it cannot be coarsened if ${\tt lvl(e)}$ is not strictly positive. In other words, an element cannot be coarsened if it is currently at the maximum size. An element also cannot be further refined if it is at the maximum allowed level of refinement ${\tt lvl_{\rm max}}$.

A criterion is set for each level of refinement. If one of the nodes belonging to an element satisfies the refinement criterion and ${\tt lvl(e) < lvl_{\rm max}}$ then it is marked for refinement. If an element no longer satisfies the refinement criterion for its current level of refinement then it is marked for coarsening. 

\paragraph{The refinement/coarsening criterion}
Consider that refinement level $l$ depends on the value of a flow variable $C^l$, and consider $C^l_{k,e}$ the value of this variable at the $k$th node of element $e$. We consider a threshold type criterion for refinement. This means that if any node $k$ belonging to element $e$ satisfies $C_{k,e} > {\tt threshold(lvl(e)+1)}$, then the element is marked for refinement. It also means that if $\forall k$ $C_{k,e} \leq {\tt threshold}(\tt lvl(e))$, then the element is marked for coarsening. 

\paragraph{Frequency of the adaptive mesh refinement procedure:}
Because AMR is not without cost, it should not be performed at every simulation time step. We define a time interval $t_{\rm amr}$ that is a multiple of the time step $\Delta t$ and allow the AMR procedure to take place at every instance that the current simulation time is a multiple of $t_{\rm amr}$. This interval should be large enough to avoid needlessly executing the AMR procedure (a needless AMR procedure would be one where no elements are coarsened or refined), and small enough to be able to adapt to substantial changes in the flow.

\begin{rem}
    {\rm An element can only be refined once per AMR iteration, even if it satisfies the criterion for a higher level of refinement. The same applies for coarsening. The level of a given element can only increase or decrease by a value of one at the most every time the AMR procedure is executed.}
\end{rem} 

\subsection{The AMR algorithm and workflow}

\begin{figure}
    \centering
    \includegraphics[scale=0.6]{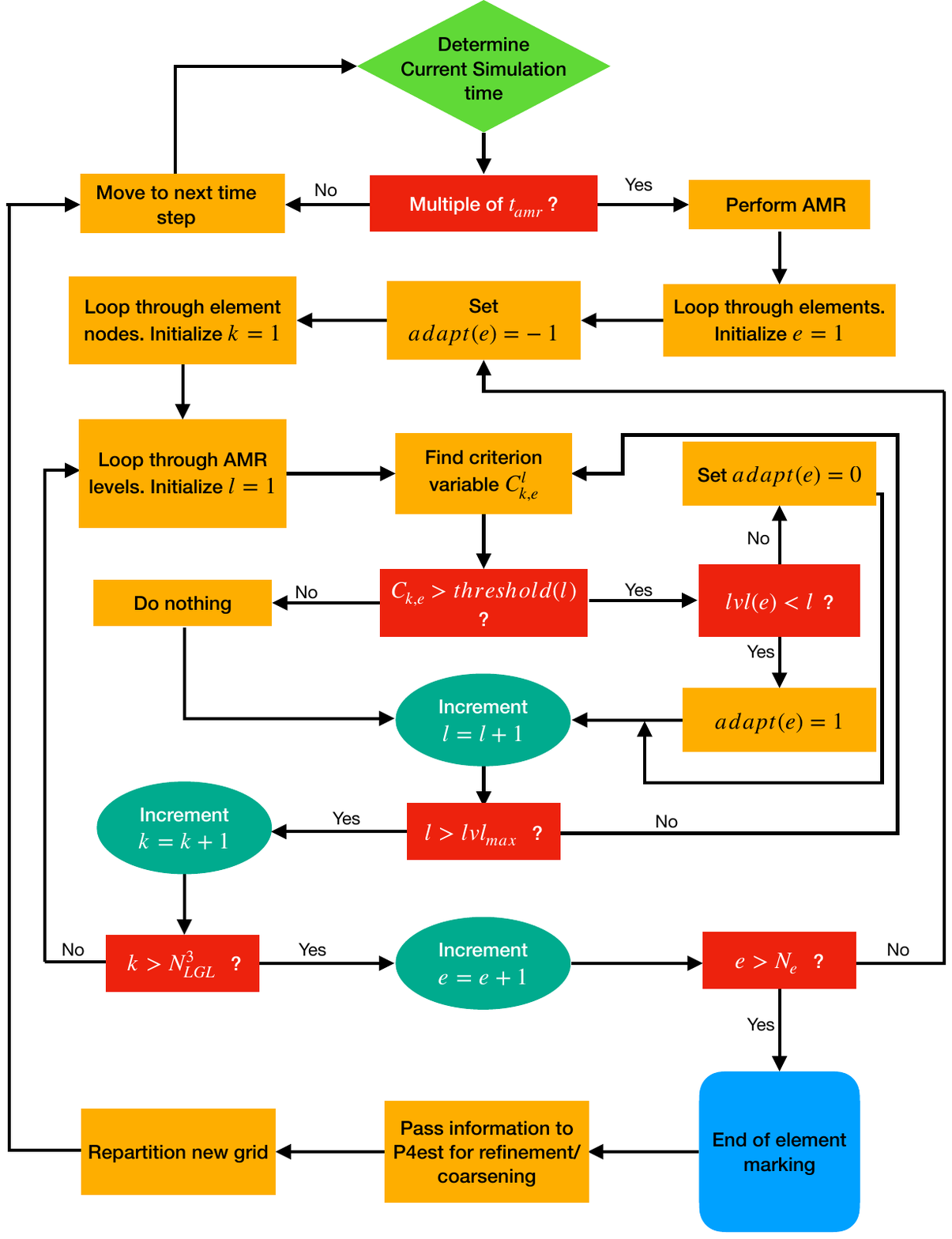}
    \caption{Workflow of the adaptive mesh refinement procedure. The start of the flow diagram is shown in the green shape labeled ``Determine Current Simulation time".}
    \label{fig:AMR_Workflow}
\end{figure}

We present the reader with an algorithmic representation of the adaptive mesh refinement process through Algorithm 3, where $t_{\rm final}$ is the final time of a given simulation, ${\tt threshold}$ is an array storing the threshold criteria for each level of refinement, $N_{LGL}$ is the number of LGL points, $N_e$ is the number of elements, and ${\tt mod}$ is the remainder operator. We also present the reader with a workflow diagram of the AMR procedure in order to illustrate the sequence of operations taking place. This diagram is shown in Figure \ref{fig:AMR_Workflow}.
{\linespread{1.1}
\begin{algorithm}[h!]\label{alg:AMR}
    \caption{Algorithm of the Adaptive Mesh Refinement Procedure}
    \begin{algorithmic}
    \For{$t =0,t_{\rm final},\Delta t$}
        \State Check if current time is a multiple of $t_{\rm amr}$.
        \If{${\tt mod}(t,t_{\rm amr}) == 0$}
            \For {$e=1,N_{e}$}
            \State ${\tt adapt}(e) = -1$
            \State elements are marked for coarsening unless they pass at least one AMR threshold. 
            \For {$k=1,N_{LGL}^3$}
            \State Compute local AMR criterion $C_{k,e}$.
            \For{$l = 1,{\tt lvl_{\rm max}}$}
            \If{$C_{k,e}^l > {\tt threshold}(kv)$}
            \If{$\tt {\tt lvl}(e) < l$} 
                \State ${\tt adapt}(e) = 1$
            \Else
                \State ${\tt adapt}(e) = 0$
            \EndIf
            \EndIf
            \EndFor
            \EndFor
            \EndFor
        \EndIf
        \State Refine and coarsen marked elements through P4est
        \State Re-partition new mesh
    \EndFor
    \end{algorithmic}
\end{algorithm}
}

\section{Simulations and Results}

The NUMA model \cite{giraldoEtAl2013} is used for all simulations. 
The initial conditions for this test are similar to those in \cite{guimondEtAl2016}. A dry tropical storm-like vortex is initialized using the following profile for azimuthal-mean tangential velocity:
\begin{equation}
    \overline{v}(r,z) = V(r)\exp\left[-\frac{z^{\sigma}}{\sigma D_1^{\sigma}}\right]\exp\left[-\left(\frac{r}{D_2}\right)^6\right],
\end{equation}
where $V$ is the surface tangential velocity, $\sigma = 2$, $D_1=5,823~$m, and $D_2=200~$km. The surface tangential velocity can be found by following the procedure described in \cite{NG03,NMS07} and integrating a specified Gaussian distribution with a vorticity peak of $1.5 \times 10^{-3}~\rm{s^{-1}}$ and maximum winds of $21.5~\rm{ms^{-1}}$ at a radius of $50~$km from the center. The vertical velocity is initially taken to be zero everywhere. 

The density and potential temperature are initialized by an iterative procedure that oscillates between satisfying the gradient wind balance and the hydrostatic balance until a specified criterion is met. This procedure is described in detail in \cite{Nolan2011}. The background state is defined by vertically interpolating the \cite{Jordan1958} mean hurricane-season sounding onto the spectral element grid.

\paragraph{Observational heating:}
The time evolution of the vortex is driven by a four dimensional source term in the energy equation. This source term represents latent heating/cooling rates in convective clouds derived from Doppler radar measurements in Hurricane Guillermo (1997). The algorithm and observations of latent heating/cooling rates is described in detail in \cite{Guimondetal2011} and its implementation in NUMA is described in \cite{Badrul2022}. Hurricane Guillermo (1997) was a rapidly intensifying TC and the observational heating data should provide an excellent testing ground for how AMR responds to the small-scale variability inherent in this complex system. As described in \cite{Guimondetal2011}, the heating is computed on a grid covering the inner core of the system out to a radius of $r=60~$km from the domain center. This grid has a resolution of $2~$km in the horizontal direction and $0.5~$km in the vertical direction. The heating observations are split into ten snapshots covering a $5.7~$hour period in intervals of $34~$minutes. The largest heating rates are present at a radius of $25-30~$km from the domain center, well within the radius of maximum winds for the initial conditions. Outside of $r=60~$km radius the heating term is zero. The heating rates are interpolated in space onto the spectral element grid. After initialization, the first heating snapshot is gradually introduced over the first $30$ minutes of the simulation, by way of a hyperbolic tangent function. The snapshots are then linearly interpolated to the next observation time over the course of the remainder of the simulation. Past $t=5.7~$hours the  heating is maintained at the last snapshot until the simulation finishes at $t=6~$hours. 

\paragraph{Boundary conditions}
At the lower boundary the surface layer model described in \cite{Surf_layer_2012} is active, while the sea-surface temperature is maintained constant at its reference value. The lateral boundaries are doubly periodic, and a Rayleigh absorbing layer is used to damp gravity waves at the domain top and is described by:
\begin{equation}
    R(z) = min\left(1,max\left(0,\gamma \sin^2\left[\frac{\pi}{2}\left(1-\frac{z_{top - z}}{z_d}\right)\right]\right)\right),
\end{equation}
where $\gamma = 1.0$, $z_{top} = 20~$km and $z_d = 4~$km. The term $q'R(z)$ is added to the right hand side of each equation where $q'=\rho'$ for Equation \eqref{eq:mass_split}, $q'=\bu$ for Equation \eqref{eq:mom_split}, and $q'=\theta'$ for Equation \eqref{eq:theta_split}.  

\paragraph{Sub-grid models}
The Smagorinsky-Lilly model \cite{smagorinsky1963,lilly1962} is used to model sub-grid scale turbulence in the horizontal direction. This model is a turbulent viscosity model and contributes to the right hand side through the divergence of the turbulent stress tensor $\nabla \cdot \mathbf{\tau}$. The turbulent stress tensor is defined as:
\begin{equation}
    \mathbf{\tau} = (-2\nu_t\mathbf{S}),
\end{equation}
where $\mathbf{S} = \frac{1}{2}(\nabla \mathbf{u} + (\nabla \mathbf{u})^T)$ is the strain rate tensor, and $\nu_t$ is turbulent viscosity and is defined as follows:
\begin{equation}
    \nu_t = (\Delta C_s)^2\sqrt{2|\mathbf{S}|^2},
\end{equation}
where $C_s$ is the constant Smagorinsky coefficient and is taken as $C_s=0.21$ for our simulations. The $\Delta$ symbol is the filter width of the Smagorinsky model and is taken to be the mean horizontal resolution of a given element.

The vertical turbulent sub-grid diffusion is modeled by the planetary boundary layer (PBL) scheme of \cite{Hong2010,Hongetal2006} and is described by:
\begin{equation}
    \frac{\partial C}{\partial t} = \frac{\partial}{\partial z}\left[ K_c \left(\frac{\partial C}{\partial z}-\gamma_c\right)-\overline{(w'c')}_h\left(\frac{z}{h^3}\right)\right],
\end{equation}
where $C$ is a prognostic variable, $K_c$ is the eddy diffusivity coefficient, $\overline{w'c'}_h$ is the flux at the inversion layer, and $\gamma_c$ is a correction to the local gradient. For a more detailed description of the PBL scheme we refer the reader to \cite{Hong2010,Hongetal2006}.

\paragraph{Computational domain} All of our simulations are run on a domain size $[-400,400]~\rm{km}\times[-400,400]~\rm{km}\times [0,20]~\rm{km}$ using continuous Galerkin spectral elements of order four.

\paragraph{Time integration} All of our simulations use an explicit third order Runge-Kutta scheme to perform time integration. The time step is maintained constant for all the tests we perform with the exception of a single AMR test with 4 levels of refinement. We use a higher vertical resolution for this test and thus require a smaller time step. 

\subsection{Time to solution comparison}

A set of 1 hour long simulations are performed to compare the time to solution of simulations with and without AMR at varying resolutions. The vertical resolution is maintained constant at $\Delta z=312~$m. A simulation using a uniform horizontal resolution of $\Delta x=\Delta y= 4~$km is used as the baseline and its time to solution $T_{4\rm{km}}$ is used to obtain a normalized time to solution $T^* = T/T_{4\rm{km}}$, where $T$ is the time to solution for a given simulation. 

\begin{figure}
    \centering
    \includegraphics[scale=0.4]{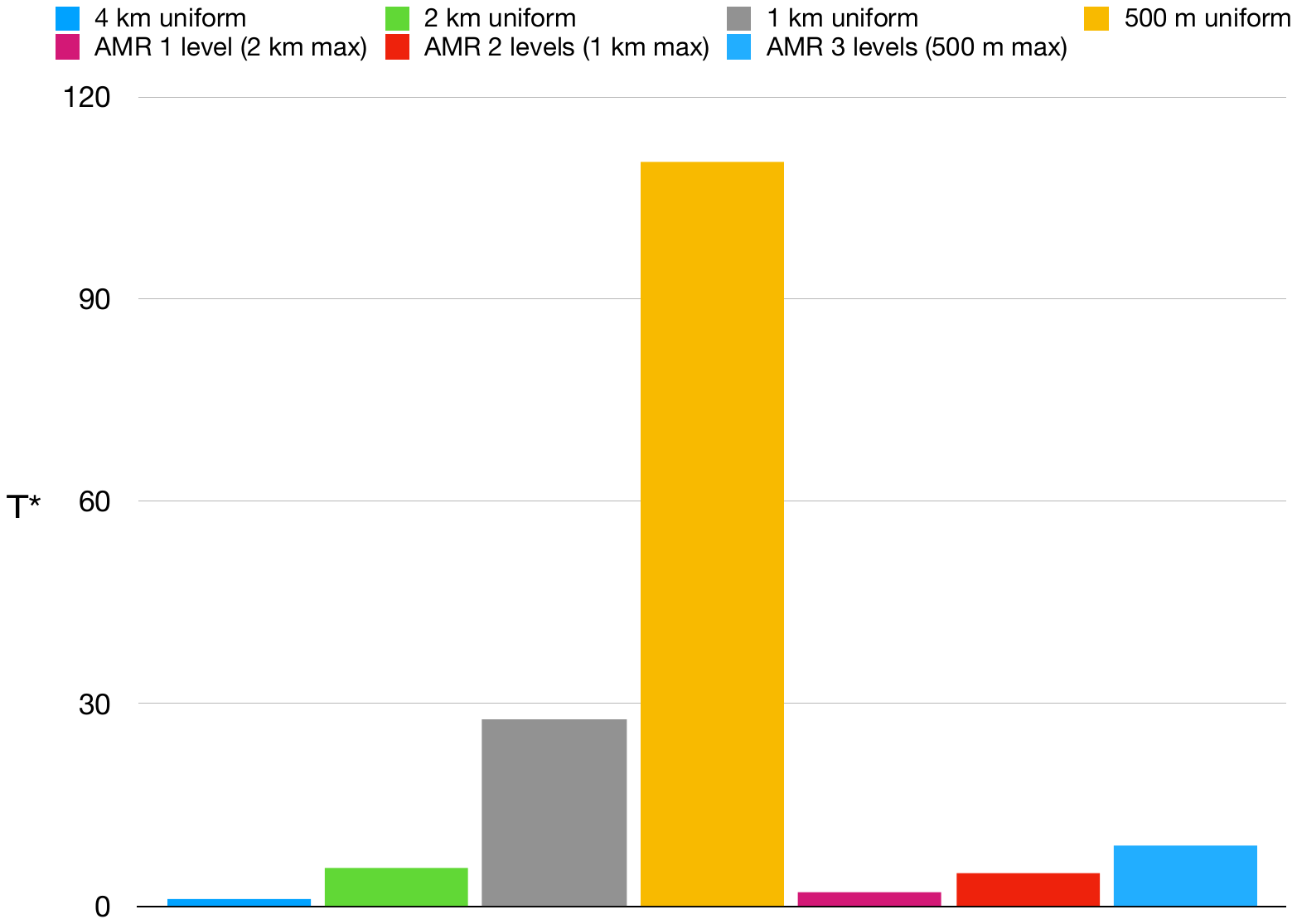}
    \caption{Time to solution comparison of uniform and AMR simulations in normalized time for the first set of tests.}
    \label{fig:AMR_times}
\end{figure}

Six simulations in total are performed for this comparison. Three constant grid spacing simulations are performed with $\Delta x= \Delta y={4,2,1}~$km, respectively. Three simulations using AMR are performed with respective minimum horizontal grid spacings of $\Delta x= \Delta y={2,1,0.5}~$km, corresponding to one, two, and three levels of refinement. The criterion for refinement in the AMR simulations is that the velocity magnitude must pass a set of predefined thresholds ${\tt threshold}_k$, where $k=1,{\tt lvl}_{\rm max}$ and ${\tt lvl}_{\rm max}$ is the maximum refinement level. For these tests we perform tests with the following thresholds: ${\tt threshold}_1=\left[7.5\right]~\rm{ms^{-1}}$, ${\tt threshold}_2=\left[7.5,15\right]~\rm{ms^{-1}}$, and ${\tt threshold}_3 = \left[7.5,15,22.5\right]~\rm{ms^{-1}}$. All simulations are performed on the same machine with the same number of cores, and we will refer to these tests as the first set in the remainder of the paper. The normalized time to solution for these tests is presented in Figure \ref{fig:AMR_times}. For the static uniform grids, the increase in horizontal resolution results in $\approx 4$ times increase in time to solution from $4~$km to $2~$km and then similarly from $2~$km to $1~$km. This is expected as the number of grid cells quadruples when doubling the horizontal resolution in a 3D simulation. We use a scaling factor of four to present an estimated time to solution for a uniform grid with a horizontal resolution $\Delta x = 500~$m on the same figure (yellow histogram). With the set of criteria being used, the time to solution for the AMR simulations with up to $1~$km horizontal resolution remain lower than a uniform $2~$km horizontal resolution simulation. Additional refinement levels results in a $\approx2$ times increase in cost for the AMR simulations with the criteria being used in these tests. The AMR simulation with a horizontal resolution of up to $500~$m is still much cheaper than a uniform $1~$km resolution simulation and barely more expensive than a uniform $2~$km resolution simulation. Compared to the estimated cost of a uniform $\Delta x =500~$m simulation the AMR simulation with a resolution of up to $\Delta x=500~$m is nearly 13 times faster.  

A second set of tests is performed to observe the effect of changing the refinement criterion on the time to solution and to see how AMR performs over the full six hour simulation period. This set of tests includes four simulations: \begin{itemize}
    \item A simulation using a uniform grid with a horizontal resolution $\Delta x= \Delta y=1~$km.
    \item A simulation with 2 levels of refinement (up to $1~$km horizontal resolution) with the same refinement criterion ${\tt threshold}_{\rm st} = {\tt threshold}_{2}$ as the previous 2-level simulation, this will be referred to as the strict 2-level simulation.
    \item A simulation with 2 levels of refinement but with a more lenient refinement criterion ${\tt threshold}_{\rm ln} = \left[2,5\right]~\rm{ms^{-1}}$, this will be referred to as the lenient 2-level simulation.
    \item A simulation with 4 levels of refinement(up to $250~$m horizontal resolution) and a vertical resolution $\Delta z=250~$m. The thresholds for the simulation are ${\tt threshold}_4 = \left[7.5,15,22.5,30\right]~\rm{ms^{-1}}$.
\end{itemize} 

\begin{rem}
    {\rm All simulations except the one with four levels of AMR maintain a vertical resolution $\Delta z=312~$m.}
\end{rem}

\begin{figure}
    \centering
    \includegraphics[scale=0.4]{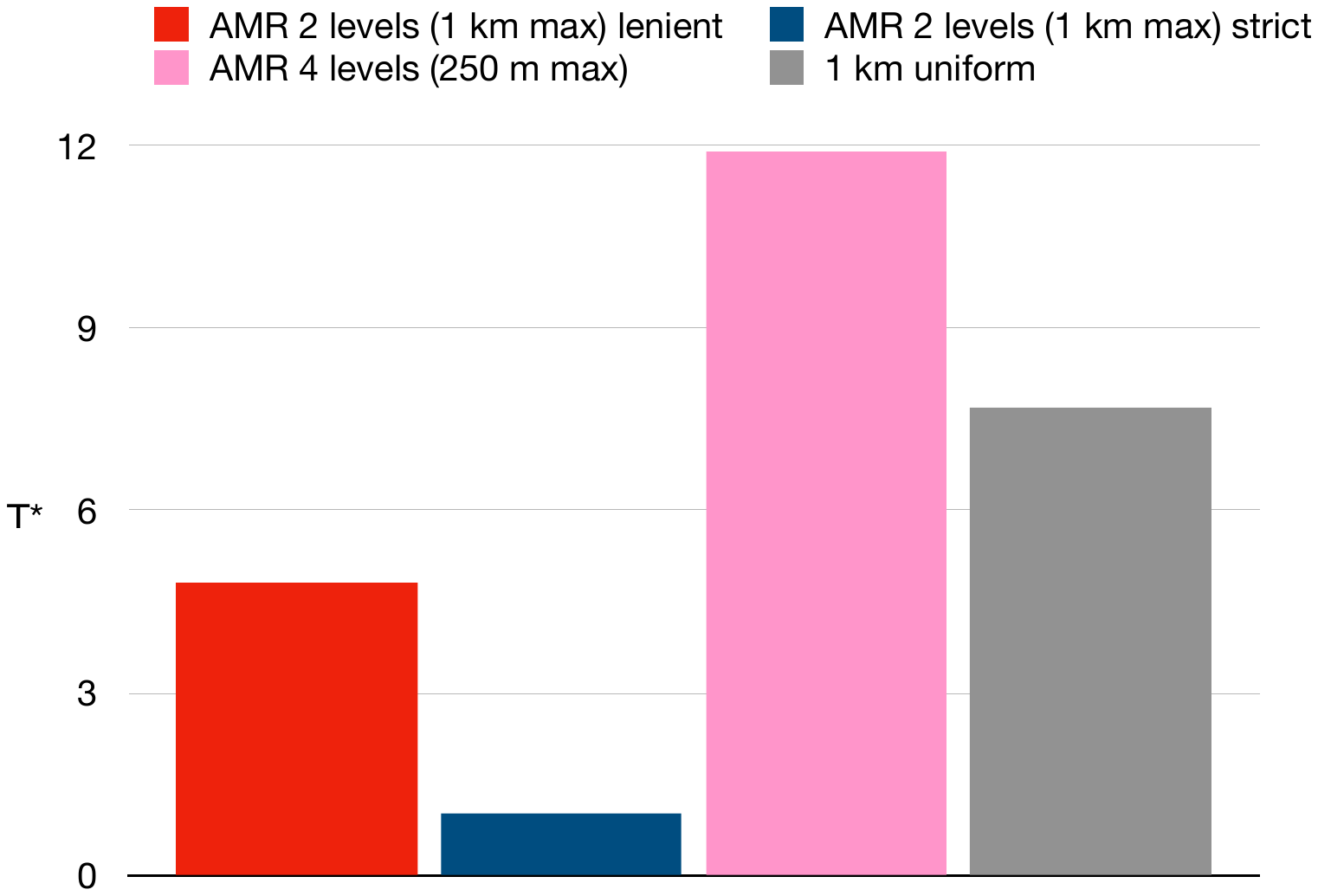}
    \caption{Time to solution comparison of uniform (static) and AMR simulations in normalized time for the second set of tests. The strict 2-level AMR uses ${\tt threshold}_{\rm st}$, while the lenient 2-level AMR uses ${\tt threshold}_{\rm ln}$}
    \label{fig:AMR_times_2}
\end{figure}

Figure \ref{fig:AMR_times_2} presents the time to solution for the second set of tests in normalized time. For these tests, the baseline is the strict 2-level AMR simulation as it is the cheapest to perform. We can see that while the two simulations using two levels of AMR remain cheaper than the uniform alternative, the more lenient criterion incurs a substantial increase in the cost of the simulation. 

\begin{figure}
    \centering
    \includegraphics[width = 0.8\textwidth]{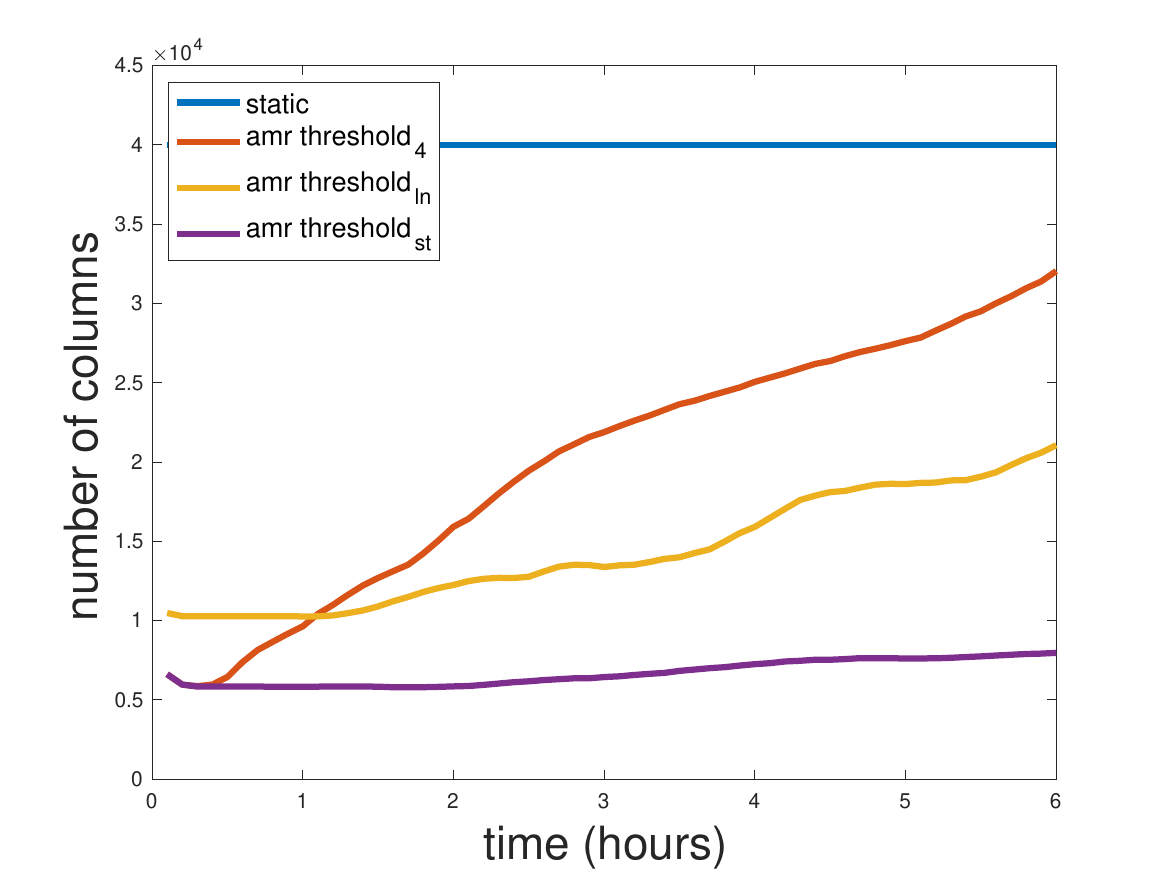}
    \caption[Evolution of number of element columns over time for AMR simulations of tropical cyclones.]{Number of element columns over time for the uniform (static) grid simulation (blue), the 4 level AMR simulation (red) with ${\tt threshold}_4 = \left[7.5,15,22.5,30\right]~\rm{ms^{-1}}$, the 2 level amr simulation with ${\tt threshold}_{\rm ln} = \left[2,5\right]~\rm{ms^{-1}}$ (yellow), and the 2 level amr simulation with ${\tt threshold}_{\rm st} = \left[7.5,15\right]~\rm{ms^{-1}}$ (purple).}
    \label{fig:Columns over time}
\end{figure}

Figure \ref{fig:Columns over time} shows the number of vertical columns over time for this set of simulations. We can see that with the stricter AMR criteria and 2 levels of refinement (purple line), the number of columns does not vary substantially over the course of the simulation. With the more lenient criteria and 2 levels of refinement (yellow line), the number of columns at the end of the simulation is double what it was at the beginning which explains the significant increase in cost. The simulation using 4 levels of AMR shows a substantial increase in the number of columns over the course of the simulation, the number of columns at the end is five times what it was at the beginning. As the hurricane intensifies more areas of high velocity magnitude appear and the mesh adapts by refining around these areas and generating more columns. While the total number of columns for the 4 level AMR simulation is less than the total number of columns for the uniform (static) simulation, it still requires a longer time to complete. The main reason for this is the increased vertical resolution which increases the total number of degrees of freedom and further restricts the time step required for stability. For the remainder of this section we will continue referring to this set of simulations. 

\subsection{Accuracy of AMR simulations}

\begin{figure}
    \centering
    \includegraphics[width = 0.8\textwidth]{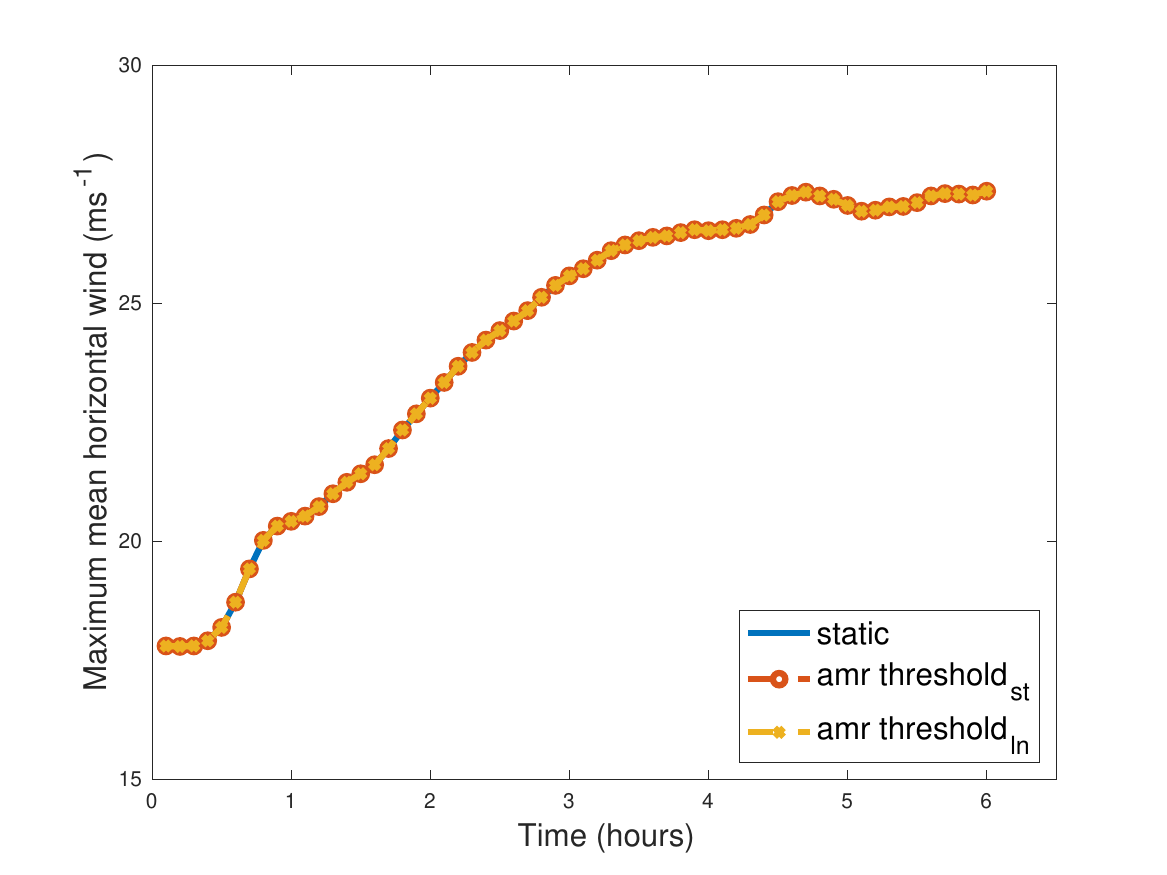}
    \caption[Maximum of horizontally averaged horizontal wind over the course of the 6 hour simulation period.]{Maximum of horizontally averaged horizontal velocity within the $[-50,50]~\rm{km^2}$ sub domain over the course of the 6 hour simulation period. The blue solid line (uniform/static simulation) overlaps perfectly with the dashed line with red circles (stricter criterion AMR simulation) and the dashed line with yellow crosses(more lenient criterion AMR simulation). }
    \label{fig:Max_mean_comparison}
\end{figure}

Figure \ref{fig:Max_mean_comparison} shows the maximum values of the horizontally averaged horizontal velocity over time for $(x,y) \in [-50,50]~\rm{km}\times[-50,50]~\rm{km}$. These values are found by finding the average value of horizontal velocity at each vertical level and then finding the maximum among these.

\begin{figure}
    \centering
    \includegraphics[width = 0.8\textwidth]{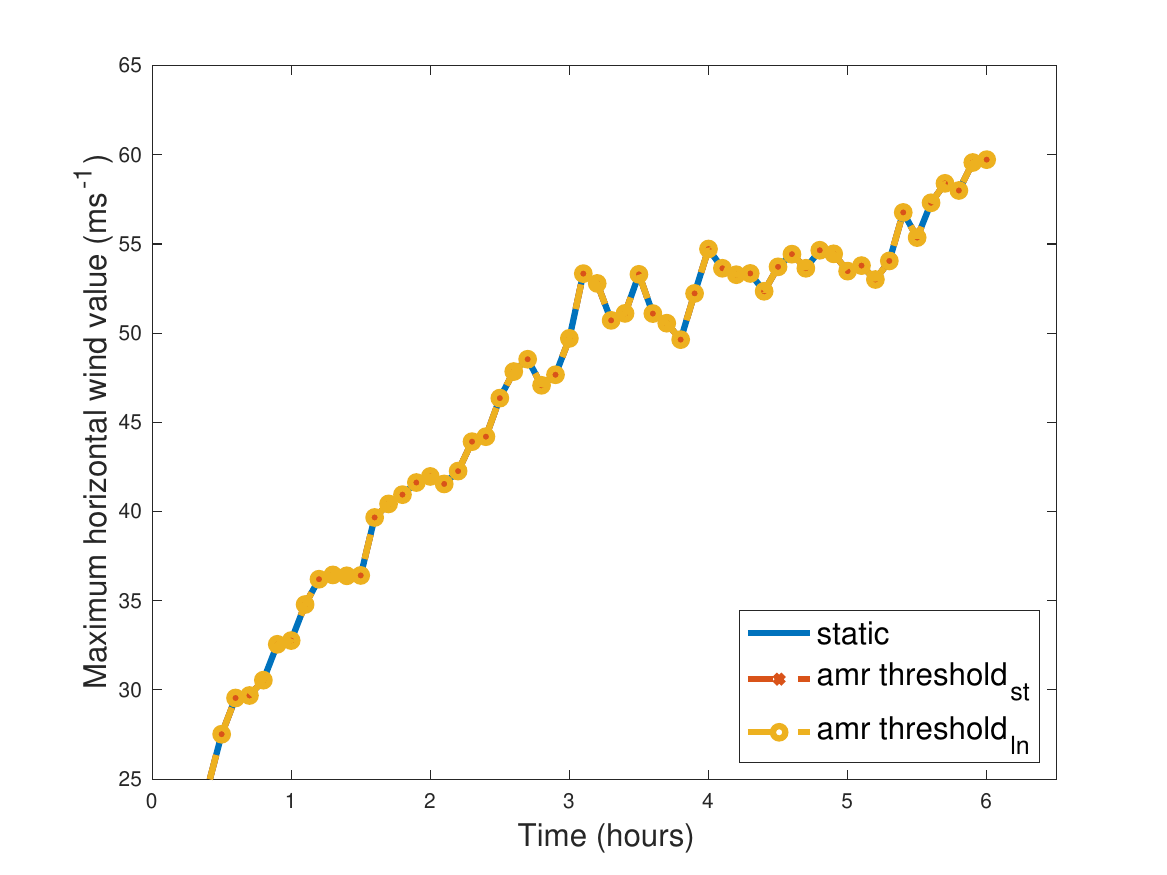}
    \caption[Maximum value of horizontal wind over the course of the 6 hour simulation period.]{Maximum value of horizontal velocity within the $[-50,50]~\rm{km}\times[-50,50]\rm{km}\times[0,20]~\rm{km}$ sub domain over the course of the 6 hour simulation period. The blue solid line (uniform/static simulation) overlaps perfectly with the dashed line with red crosses (stricter criterion AMR simulation) and the dashed line with yellow circles(more lenient criterion AMR simulation).}
    \label{fig:Max_wind_comparison}
\end{figure}

This horizontal sub-domain is chosen as it should allow for focusing on where the observational heating takes place. Figure \ref{fig:Max_wind_comparison} shows the values of maximum horizontal velocity over time. Both figures compare these values for the uniform simulation and the two 2-level AMR simulations. Both figures show an essentially perfect overlap for these quantities and demonstrate the AMR's ability to capture the intensification of the storm even with relatively strict criteria.

\begin{figure}
\centering
\includegraphics[width=\textwidth,height=510pt]{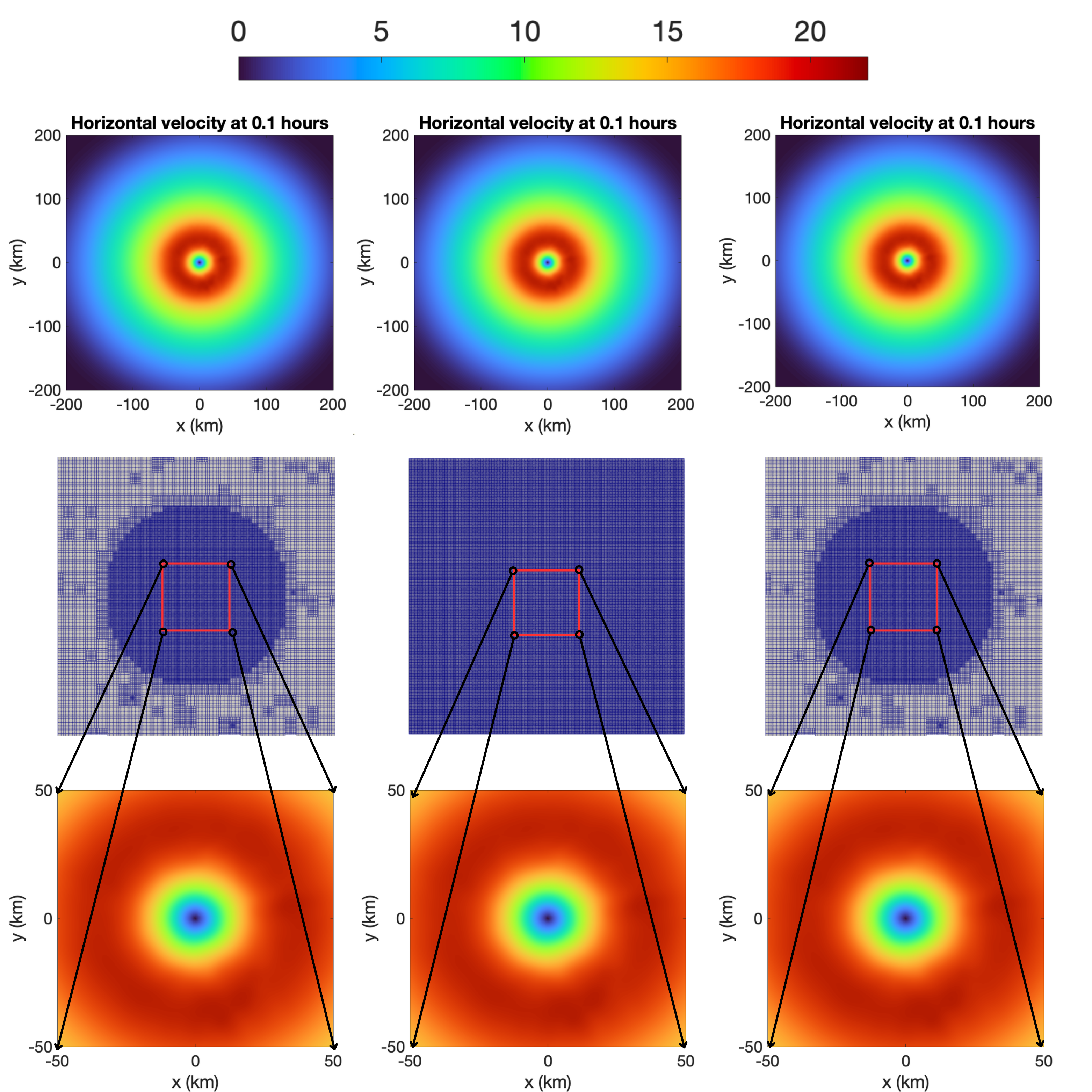}
      
    \caption[Horizontal winds taken at height $z=1000~$m above sea level at different $t=0.1~$hours.]{Horizontal winds taken at height $z=1000~$m above sea level at $t=0.1~$hours. The left column presents results for a simulation with 2 levels of AMR and a maximum horizontal resolution of $1~$km. The center column presents the results for a uniform grid with $1~$km constant horizontal resolution. The right column presents the results for a simulation with a maximum of 4 levels of AMR. At this early stage of the simulation only two levels of refinement can be seen even on the 4-level capable simulation as the criteria for higher refinement aren't met yet.}
    \label{fig:horizontal slices}
\end{figure}

\begin{figure}
\centering
    \includegraphics[width=\textwidth,height=510pt]{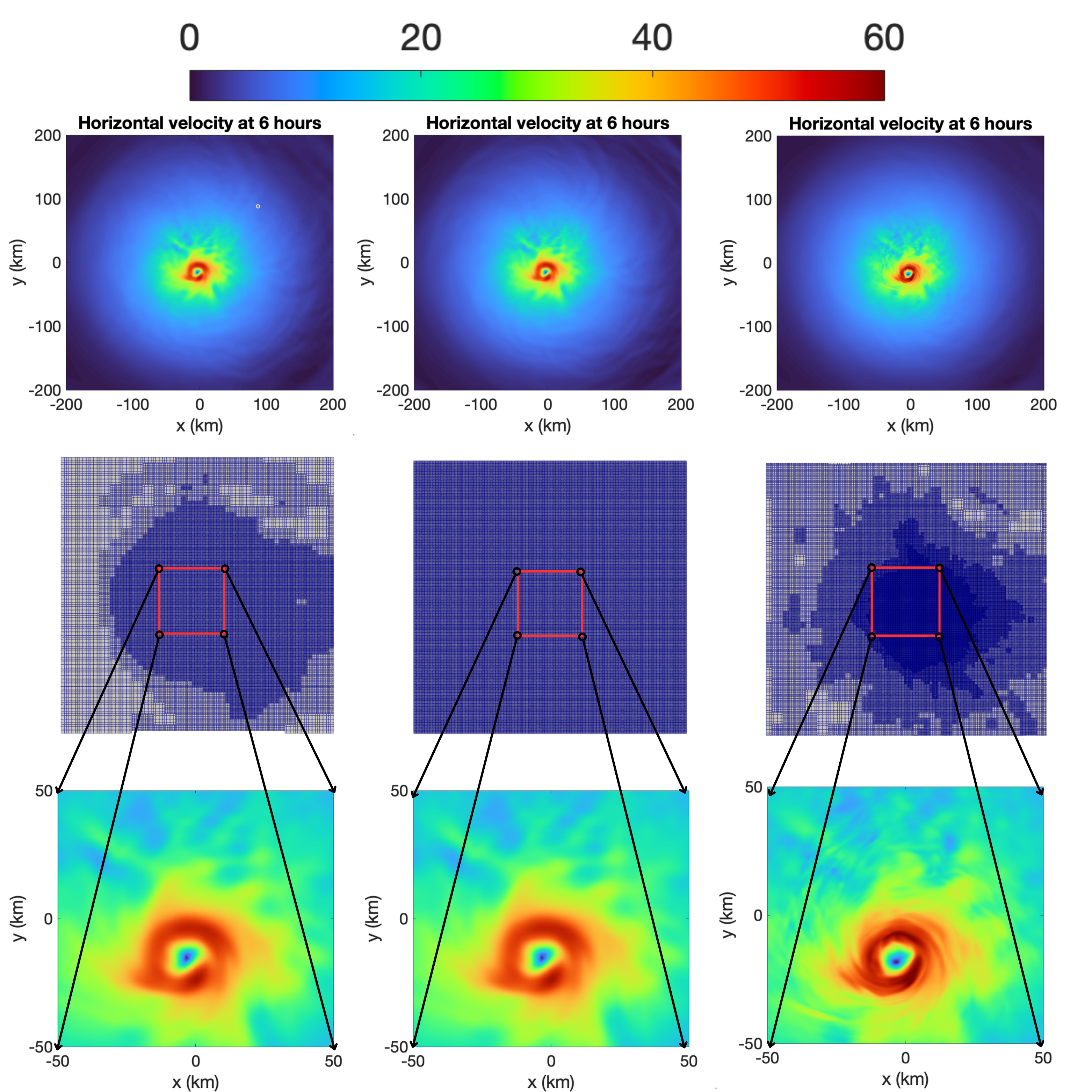}
      
    \caption[Horizontal winds taken at height $z=1000~$m above sea level at $t=6~$hours.]{Horizontal winds taken at height $z=1000~$m above sea level at $t=6~$hours. The left column presents results for a simulation with 2 levels of AMR and a maximum horizontal resolution of $1~$km. The center column presents the results for a uniform grid with $1~$km constant horizontal resolution. The right column presents the results for a simulation with 4 levels of AMR and a maximum horizontal resolution of $250~$m.All 4 levels of refinement are visible on the grid in the rightmost column.}
    \label{fig:horizontal slices_6_hours}
\end{figure}

\begin{figure}
\centering
   \includegraphics[width=\textwidth,height=510pt]{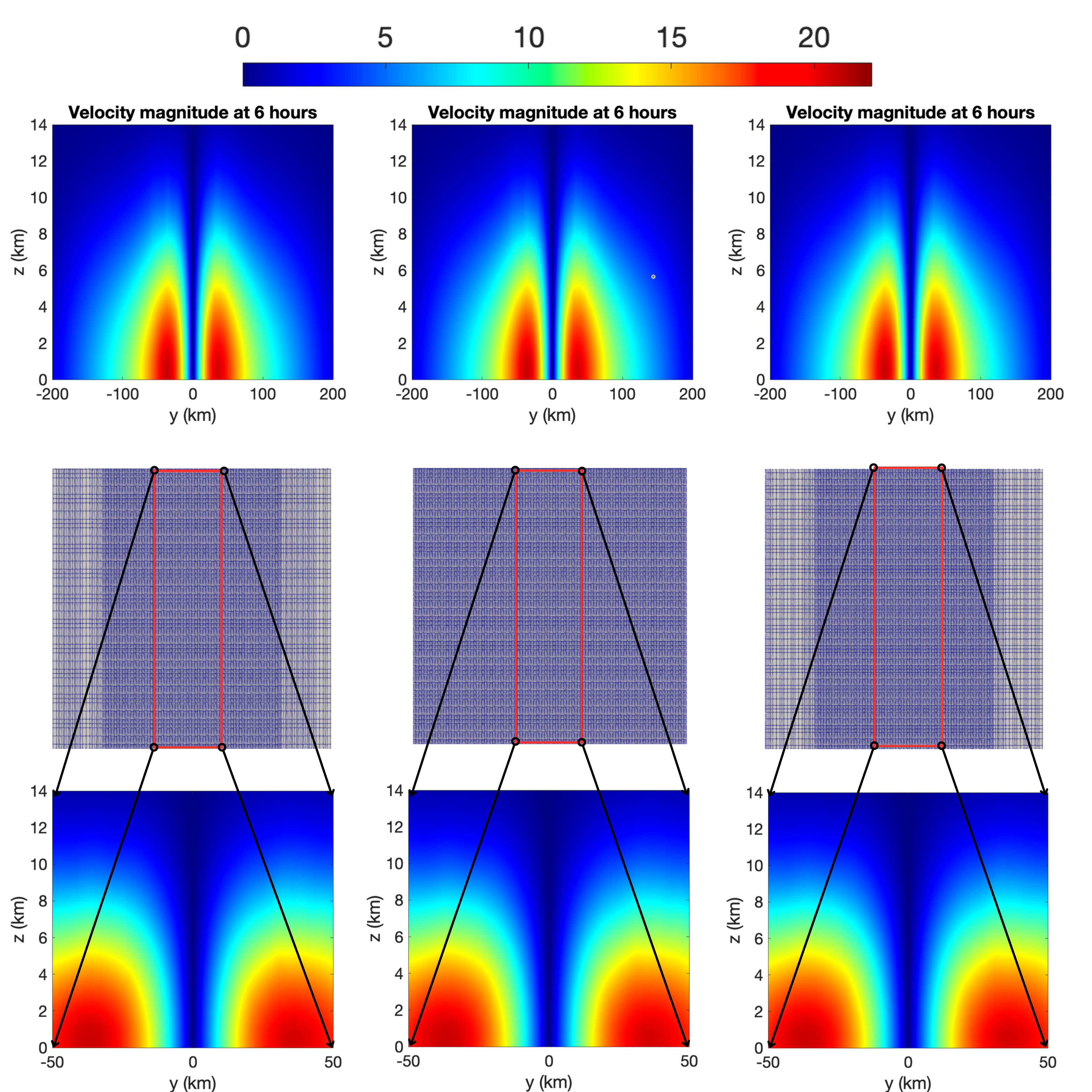}
    \caption[Velocity magnitude taken at $x=0~$m at $t=0.1~$hours.]{Velocity magnitude taken at $x=0~$m at $t=0.1~$hours. The left column presents results for a simulation with 2 levels of AMR and a maximum horizontal resolution of $1~$km. The center column presents the results for a uniform grid with $1~$km constant horizontal resolution. The right column presents the results for a simulation with 4 levels of AMR and a maximum horizontal resolution of $250~$m. At this early stage of the simulation only two levels of refinement can be seen even on the 4-level capable simulation as the criteria for higher refinement aren't met yet.}
    \label{fig:vertical slices}
\end{figure}

\begin{figure}
\centering
    \includegraphics[width=\textwidth,height=510pt]{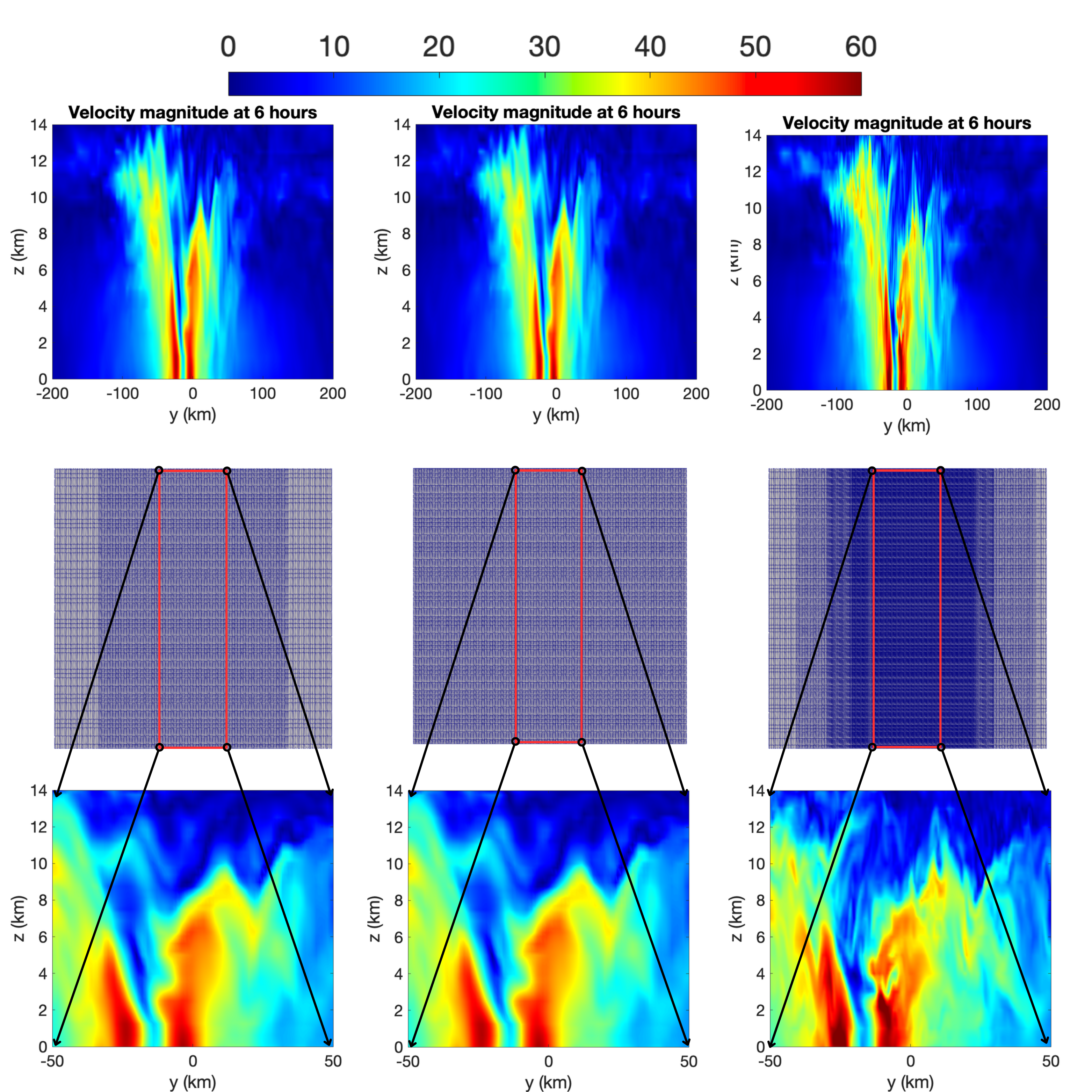}
    \caption[Velocity magnitude taken at $x=0~$m at $t=6~$hours.]{Velocity magnitude taken at $x=0~$m at $t=6~$hours. The left column presents results for a simulation with 2 levels of AMR and a maximum horizontal resolution of $1~$km. The center column presents the results for a uniform grid with $1~$km constant horizontal resolution. The right column presents the results for a simulation with 4 levels of AMR and a maximum horizontal resolution of $250~$m. All 4 levels of refinement are visible on the grid in the rightmost column.}
    \label{fig:vertical slices 6 hours}
\end{figure}

\begin{figure}
\centering
    \includegraphics[width=\textwidth,height=510pt]{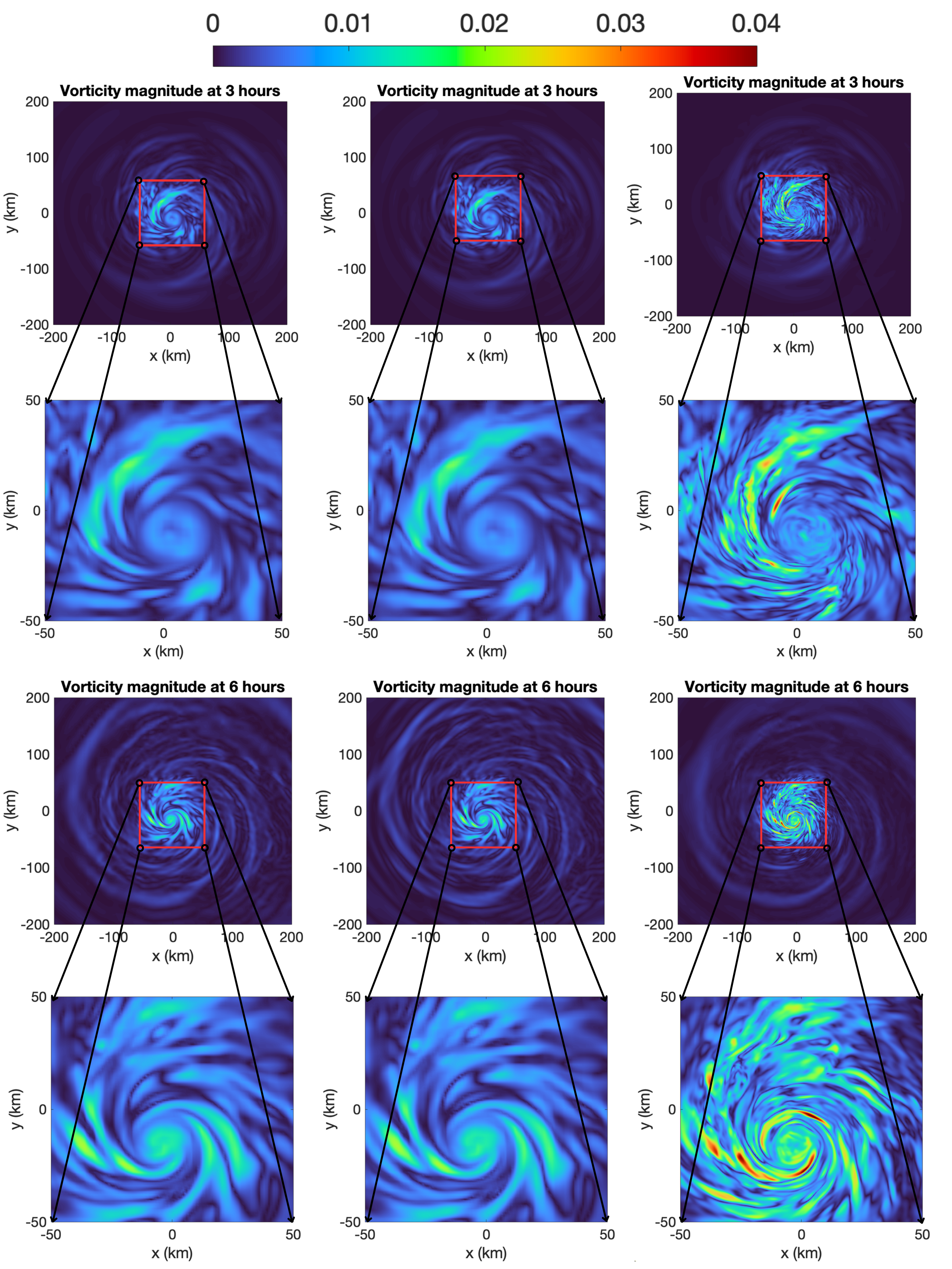}
    \caption[Vorticity magnitude taken at height $z=1000~$m above sea level at different stages of the tropical cyclone simulation.]{Vorticity magnitude taken at height $z=1000~$m above sea level at $t=3~$hours (top two rows) and $t=6~$hours (bottom two rows). The left column presents results for a simulation with 2 levels of AMR and a maximum horizontal resolution of $1~$km. The center column presents the results for a uniform grid with $1~$km constant horizontal resolution. The right column presents the results for a simulation with 4 levels of AMR and a maximum horizontal resolution of $250~$m.}
    \label{fig:vorticity slices}
\end{figure}

Figure \ref{fig:horizontal slices} shows the horizontal velocity of the storm at $t=0.1$~hours for a horizontal slice taken at $z=1000~$m. We can see that the left column (2-level AMR with strict criterion), middle column (uniform grid) and right column (4-level AMR) are identical. At this early stage of the simulation the 4-level AMR simulation behaves identically to the 2-level AMR simulation because the criteria for the third and fourth levels of refinement aren't met yet. Figure \ref{fig:horizontal slices_6_hours} is similar to Figure \ref{fig:horizontal slices} but corresponds to the end of the simulations at $t=6~$hours. The left and middle columns (2-level AMR and uniform grid respectively) are still identical even at the end of the simulation. The right column (4-level AMR) allows us to see the additional details and structures that can be captured with additional refinement thanks to criteria for all 4 levels of refinement being met in the domain center. Figure \ref{fig:vertical slices} shows the velocity magnitude of the storm at $t=0.1~$hours for a vertical slice taken at $x=0~$m. Once again, all the columns are identical at this stage of the simulation. Figure \ref{fig:vertical slices 6 hours} is similar to Figure \ref{fig:vertical slices} but corresponds to the end of the simulations at $t=6~$hours. The left and middle columns are still identical and demonstrate the ability of AMR to obtain high fidelity results at a fraction of the cost of a simulation using a uniform grid. Figure \ref{fig:vorticity slices} presents the vorticity magnitude of the storm at $t=3~$hours (top two rows) and $t=6~$hours (bottom two rows) for a horizontal slice at $z=1000~$m. Even for this derived quantity the left and center columns are still identical. The right column showcases what additional resolution has to offer in terms of resolving turbulent structures. The simulations with resolutions of up to $250~$m of horizontal resolution show an increase in both vorticity and velocity compared to their lower resolution counterparts. This reflects the role of the additional resolution in capturing the intensification of the storm. Vorticity is especially important when considering features like strong updrafts and convective towers. Pushing this resolution even further could allow for studies of how these features affect TC intensification and RI through numerical experiments.

\begin{figure}
    \centering
    \includegraphics[width=\textwidth]{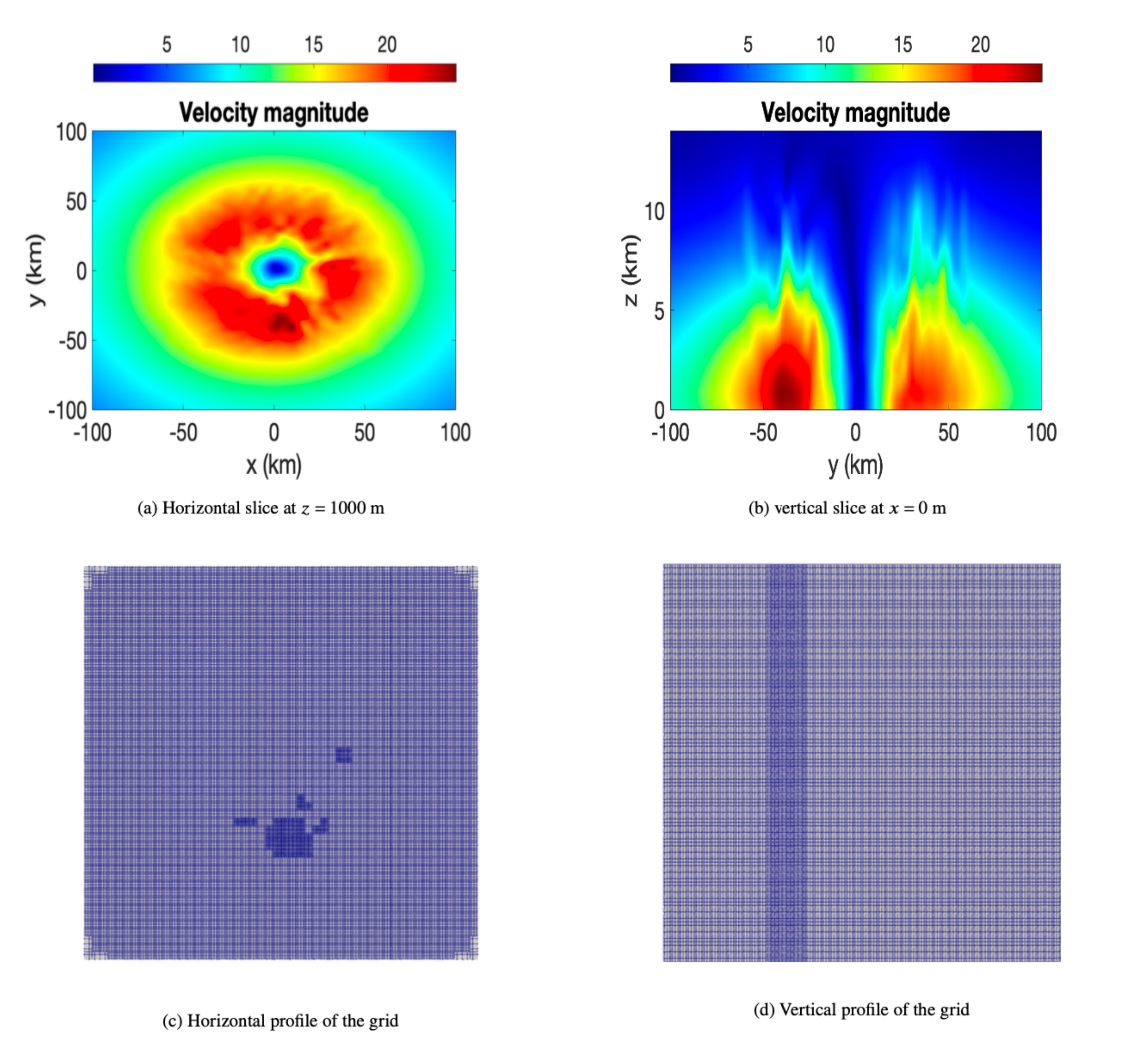}
    \caption[Velocity magnitude and AMR grid at the first instance of refinement to $500~$m of horizontal resolution.]{Velocity magnitude and grid at the first instance of reaching the third level of refinement for the 4 level AMR simulation. a) Velocity magnitude at $z=1000~$m and $t=0.3~$hours. b) Velocity magnitude at $x=0~$m and $t=0.3~$hours. c) Horizontal grid at $z=1000~m$, $t=0.3~$hours, and for $(x,y)\in[-100,100]~\rm{km}\times[-100,100]~\rm{km}$. d) Vertical grid at $x=0~$m, $t=0.3~$hours, and for $(y,z) \in [-100,100]~\rm{km}\times[-0,20]~\rm{km}$. Note that the lowest resolution visible in c) is $2~$km (seen in the corners) and the highest is $500~$m. The baseline $4~$km resolution is not visible as it present farther away from the domain center. In d) the lowest visible horizontal resolution is $1~$km and the highest is $500~$m. The lower resolutions are present in other areas of the simulated domain not pictured here. Areas with a velocity magnitude larger than $22.5~\rm{ms^{-1}}$ trigger the criterion for the third level of refinement and cause grid to reach $500~$m of horizontal resolution wherever the criterion is met. As this refinement is done on a column basis, the entire vertical column is refined as shown in d).}
    \label{fig:first_500m_refinement}
\end{figure}

\begin{figure}
    \centering
    \includegraphics[width=\textwidth]{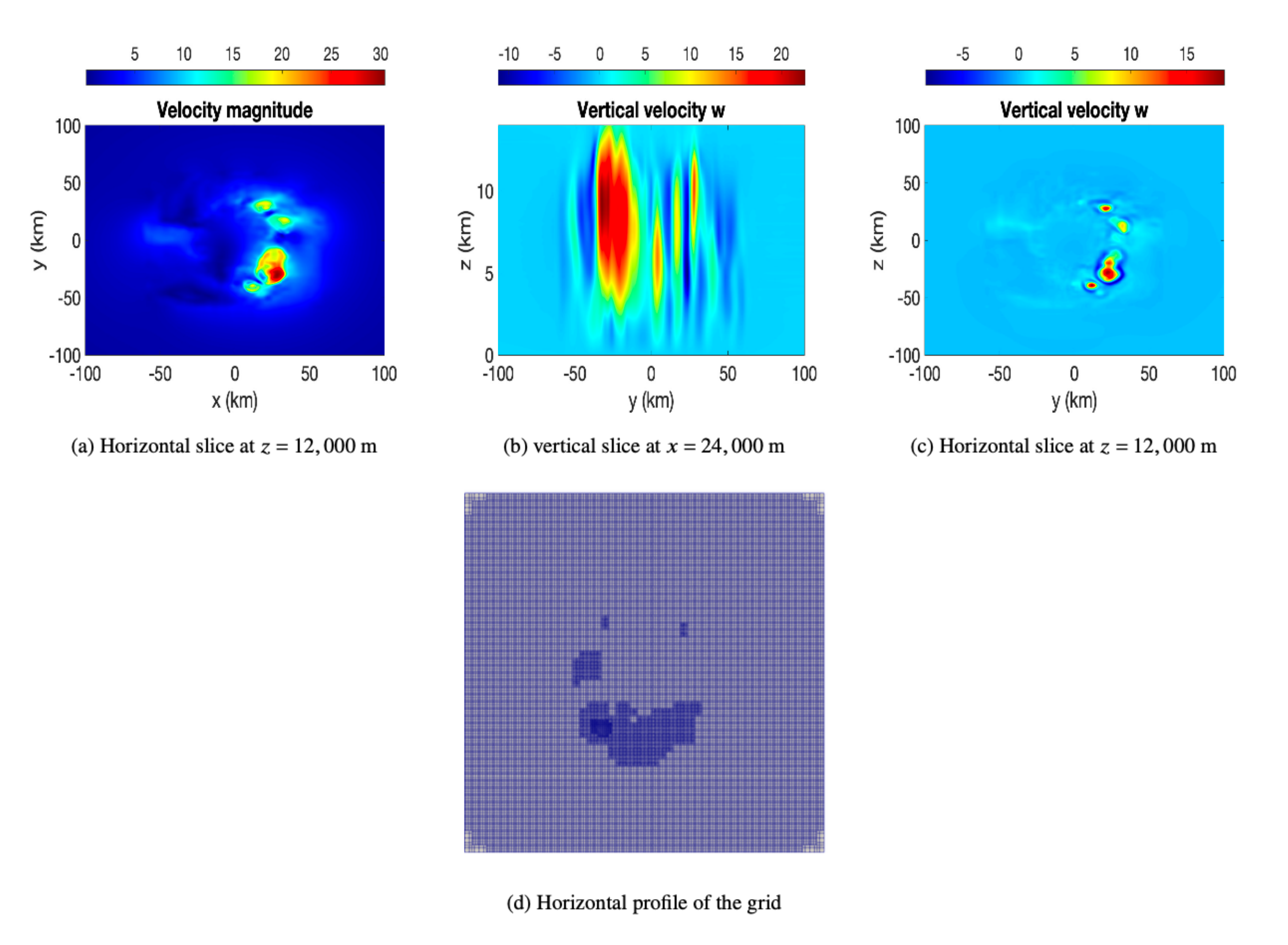}
    \caption[AMR capturing a powerful updraft: velocity magnitude, vertical velocity and grid at the first instance of AMR refining to $250~$m of horizontal resolution.]{Velocity magnitude and grid at the first instance of reaching the third level of refinement for the 4 level AMR simulation. a) Velocity magnitude at $z=12,000~$m and $t=0.4~$hours. b) Velocity magnitude at $x=24,000~$m and $t=0.4~$hours. c) Vertical velocity at $z=12,000~$m and $t=0.4~$hours. d) Horizontal grid at $z=12,000~m$, t=$0.4~$hours and for $(x,y)\in[-100,100]~\rm{km}\times[-100,100]~\rm{km}$. The lowest resolution visible in c) is $2~$km (seen in the corners) and the highest is $250~$m. In d) the lowest visible horizontal resolution is $1~$km and the highest is $250~$m. A substantial updraft indicated by the elevated vertical velocity shown in b) and c) allows the velocity magnitude to pass the $30~\rm{ms^{-1}}$ threshold required for the 4th level of AMR. The AMR is effectively refining around this updraft.}
    \label{fig:first_250m_refinement}
\end{figure}

We now focus on the simulation using 4 levels of refinement. Figure \ref{fig:first_500m_refinement} shows the first instance we detect of the grid refining to a horizontal resolution $\Delta x = \Delta y = 500~$m. This refinement takes place at $t\approx0.3~$hours, near the beginning of the simulation. The top two plots show areas of higher velocity magnitude (dark red shades) which corresponds to the criterion for refining to $500~$m being met and the bottom two plots of the figure show how the grid responds to the intensification. Figure \ref{fig:first_250m_refinement} shows the first instance we detect of the grid refining to a horizontal resolution $\Delta x=\Delta y=250~$m. This refinement occurs at $t\approx0.4~$hours, still near the beginning of the simulation. As shown in the top left plot, high values of velocity magnitude trigger the refinement process causing the grid to adapt (middle bottom plot). At this time the high value of velocity magnitude is associated with the occurrence of a strong updraft, as evidenced by the high vertical velocity values shown in the top-middle and top-right plots of the figure. 

\begin{figure}
    \centering
    \includegraphics[width=0.65\textwidth]{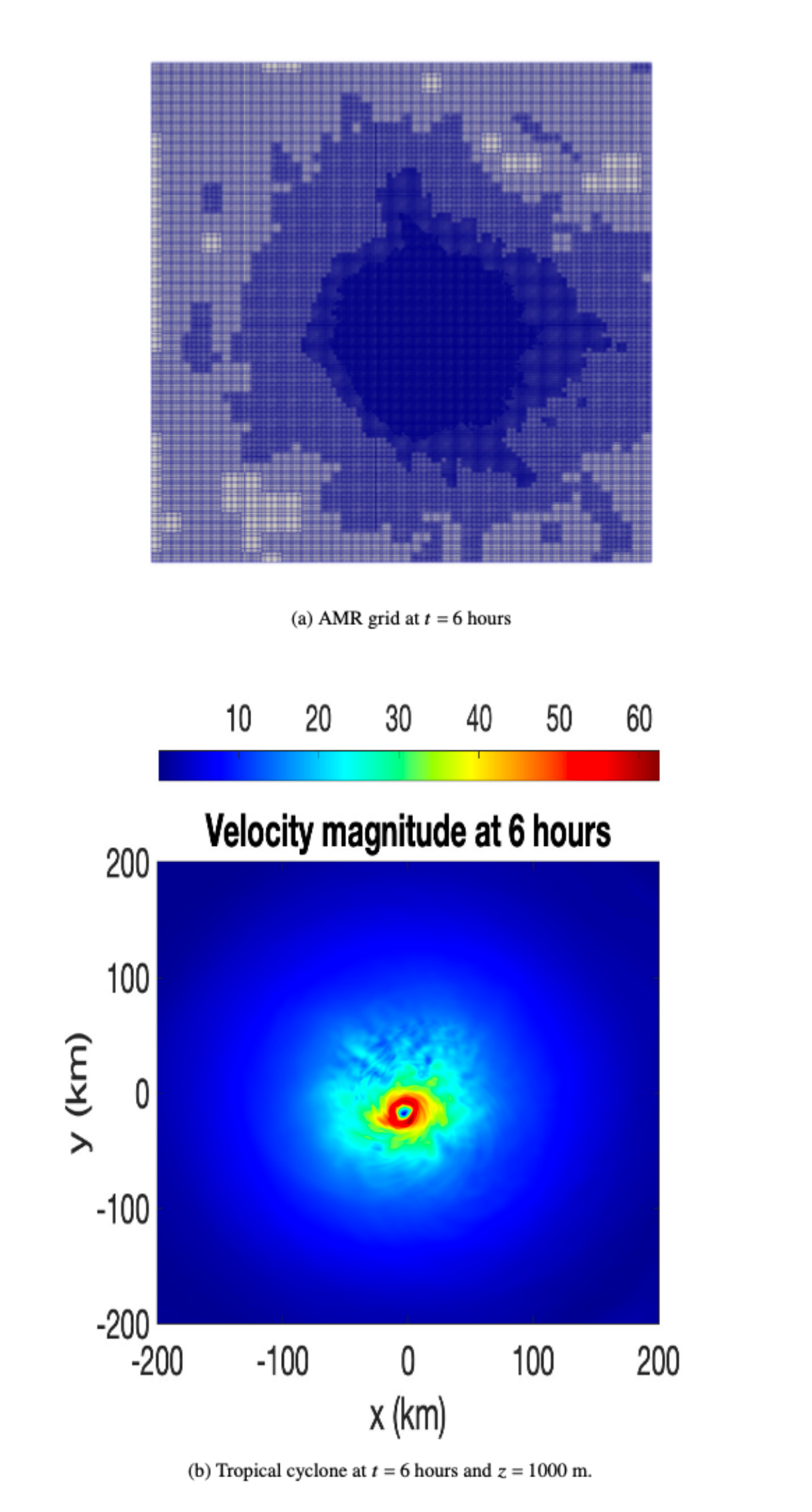}
    \caption[A 4 level AMR grid at the end of a six hour tropical cyclone simulation.]{Example of 4-level refinement at $t =6~$hours. The criterion used is ${\tt threshold}_4 = \left[7.5,15,22.5,30\right]~\rm{ms^{-1}}$ on velocity magnitude. a) Horizontal cross section of the AMR grid. b) Velocity magnitude at $t=6~$hours and $z=1000~$m.}
    \label{fig:4-level grid end}
\end{figure}

Figure \ref{fig:4-level grid end} shows the 4 level AMR simulation at $t=6~$hours and at the $z=1000~$m horizontal plane within the boundaries of $(x,y) \in[-200,200]~\rm{km} \times [-200,200]~\rm{km}$. All the levels of the AMR grid are visible and as we would expect, the grid is finer near the storm center than it is farther away. 

\begin{figure}
    \centering
    \includegraphics[width=0.7\textwidth]{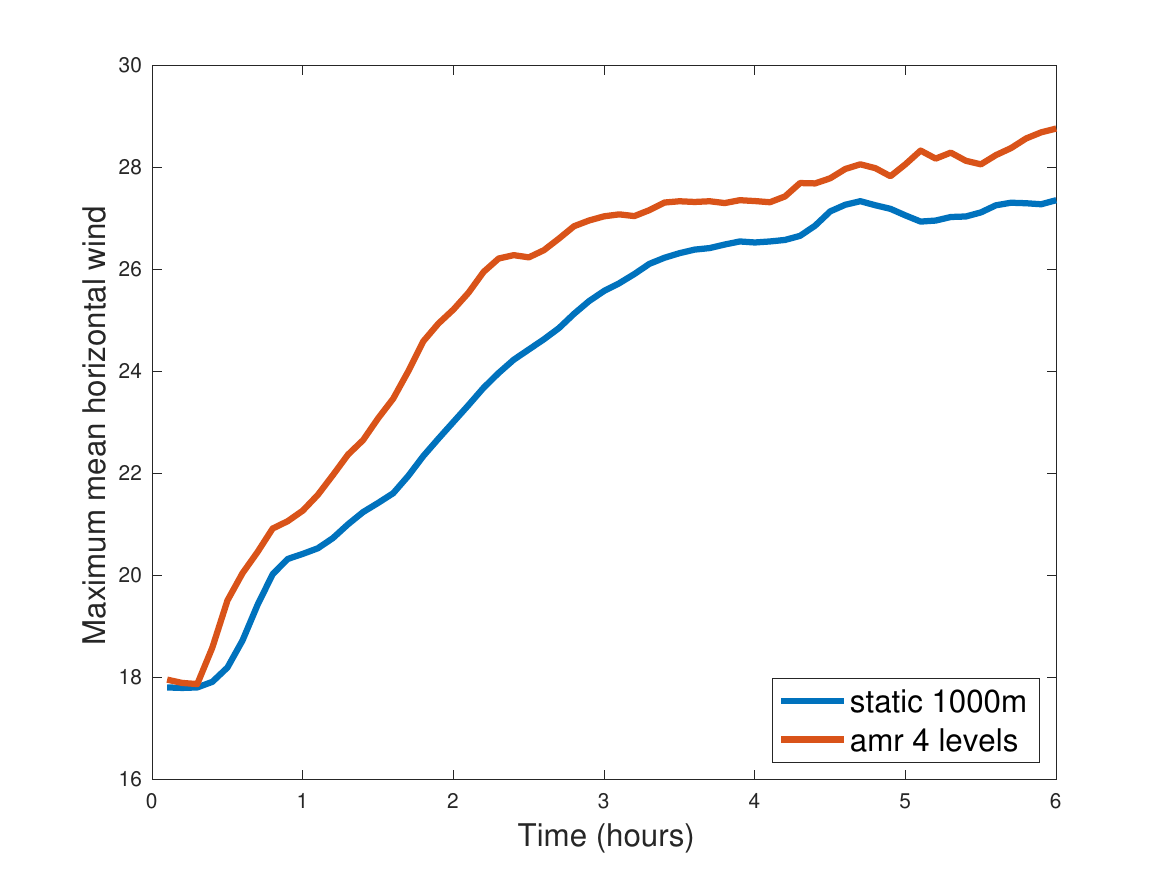}
    \caption{Maximum of horizontally averaged horizontal wind within the $[-50,50]~\rm{km^2}$ sub domain over the course of the 6 hour simulation period for the uniform $1~\rm{km}$ resolution simulation (blue) and the 4 level AMR simulation (red).}
    \label{fig:amr_vs_static_horz_avg}
\end{figure}

\begin{figure}
    \centering
    \includegraphics[width=0.7\textwidth]{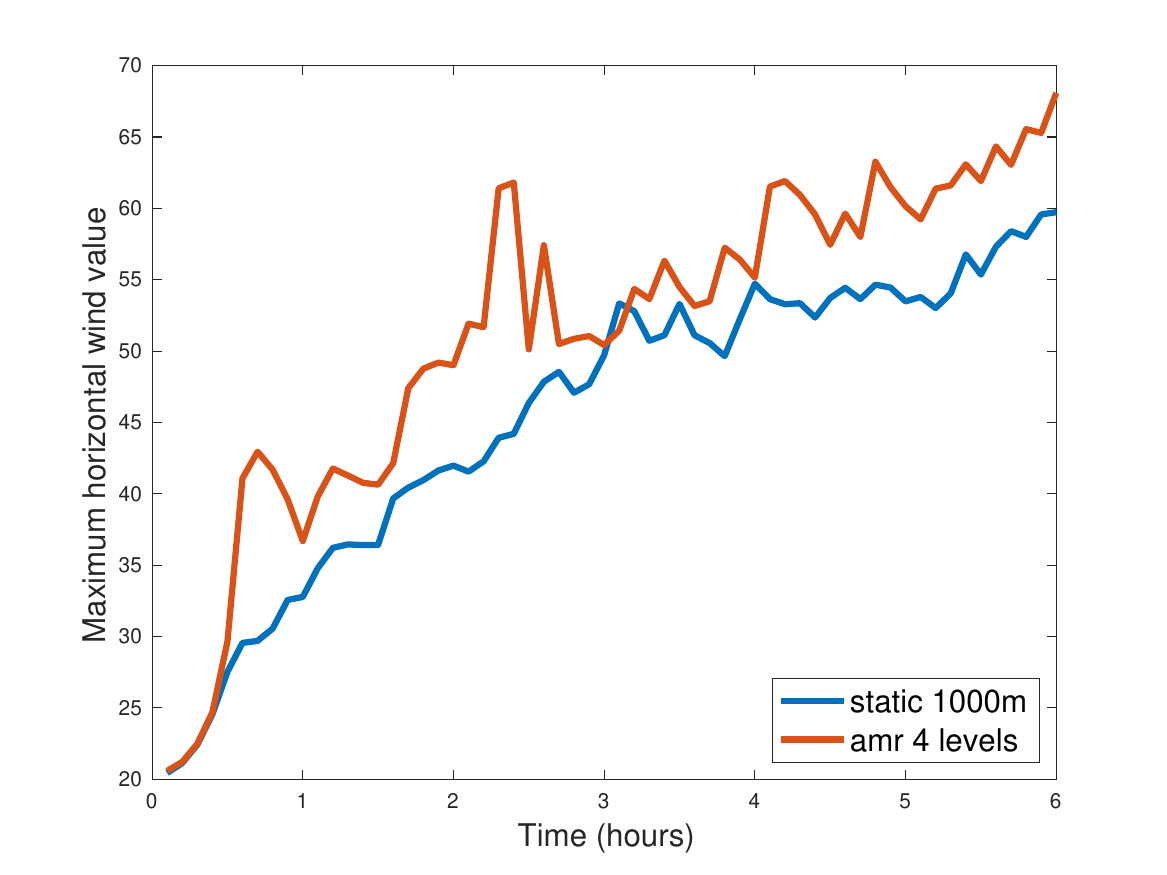}
    \caption{Maximum value of horizontal wind within the $[-50,50]~\rm{km^2}$ sub domain over the course of the 6 hour simulation period for the uniform $1~km$ resolution simulation (blue) and the 4 level AMR simulation (red).}
    \label{fig:amr_vs_static_horz_max}
\end{figure}

Figures \ref{fig:amr_vs_static_horz_avg} and \ref{fig:amr_vs_static_horz_max} show the maximum horizontally averaged horizontal wind over time, and the maximum value of horizontal wind over time, respectively. Both figures show a close overlap for the earliest stages of the respective simulations (up to $t=0.3~$hours) but begin to separate at $t=0.4~$hours for Figure \ref{fig:amr_vs_static_horz_avg} and $t=0.5~$hours for Figure \ref{fig:amr_vs_static_horz_max}. These times correspond to instants the AMR simulation first refines to $\Delta x = \Delta y =500~$m ($t=0.4~$hours) and first refines to $\Delta x = \Delta y=250~$m ($t=0.5~$hours). Both figures show that the 4 level AMR simulation produces a more intense storm with the difference in intensity generally being between (5 and 10 percent). 
Both storms are responding to exactly the same forcing from the observational heating and Figures \ref{fig:Max_mean_comparison} and \ref{fig:Max_wind_comparison} show that for the same maximum resolutions the AMR and uniform simulations are able to produce the same results. As such, the difference in storm intensity being observed here is a result of the difference in maximum resolution between the two simulations. 

\section{Discussion and Conclusions}

The efficiency of AMR depends very closely on the choice of criterion chosen for refinement. A criterion that is too strict and hard to fulfill can result in insufficient refinement and lead to less accurate simulations. A criterion that is too lenient and easy to fulfill may result in less time savings and perhaps unnecessary costs. This work does not offer an in depth study of different criteria for AMR, but we do demonstrate how it can affect the efficiency of a simulation. Deciding on a criterion requires knowledge of specific features that the AMR is meant to capture, velocity magnitude was chosen for the simulations here as it allows for AMR to detect regions where intensification takes place. However, it could be argued that this criterion can only be activated after intensification has already begun taking place and it might not be able to capture the onset of intensification. Potential vorticity has been studied by \cite{hendricksEtAl2016} as a potential criterion for TC AMR, which showed promise in a simple shallow water model. Perhaps a criterion that combines multiple flow variables could be useful for future AMR simulations of TCs.

This work presented AMR as a tool for accelerating the simulation of TCs. The algorithm used to adapt the grid was presented and its effectiveness demonstrated through a suite of tests on a dry rapidly intensifying TC case. We have demonstrated that the results of simulations using uniform grids can be accurately replicated through the use of AMR. We have also shown that AMR allows for improving the time efficiency of rapidly intensifying TC simulations. Though dependent on the adaptivity criterion, we have shown that AMR simulations can be performed at a fraction of the time it takes to perform their uniform grid counterparts. We have also shown that it is possible to perform higher resolution AMR simulations in comparable and sometimes even less time than it would take to perform a lower resolution uniform grid simulation.

\clearpage
%%%%%%%%%%%%%%%%%%%%%%%%%%%%%%%%%%%%%%%%%%%%%%%%%%%%%%%%%%%%%%%%%%%%%
% ACKNOWLEDGMENTS
%%%%%%%%%%%%%%%%%%%%%%%%%%%%%%%%%%%%%%%%%%%%%%%%%%%%%%%%%%%%%%%%%%%%%
\paragraph{Acknowledgments}
Stephen R. Guimond and Simone Marras acknowledge the partial support by the National Science Foundation through Grants PD-2121366 and PD-2121367 which supported the work by graduate student Yassine Tissaoui. Francis X. Giraldo gratefully acknowledges the support of the Office of Naval Research under grant \# N0001419WX00721 and National Science Foundation under grant AGS-1835881. The authors gratefully acknowledge the contributions of several persons who, over many years, contributed to the AMR capability in NUMA. A brief list includes Andreas M\"{u}ller, Michal Kopera, and Jeremy Kozdon. Discussions with Lucas Wilcox on P4est also helped to get this work over the finish line.
%  Keep acknowledgments (note correct spelling: no ``e'' between the ``g'' and
% ``m'') as brief as possible. In general, acknowledge only direct help in
%  writing or research. Financial support (e.g., grant numbers) for the work done, 
%  for an author, or for the laboratory where the work was performed must be 
%  acknowledged here rather than as footnotes to the title or to an author's name.
%  Contribution numbers (if the work has been published by the author's institution 
%  or organization) should be placed in the acknowledgments rather than as 
%  footnotes to the title or to an author's name.

%%%%%%%%%%%%%%%%%%%%%%%%%%%%%%%%%%%%%%%%%%%%%%%%%%%%%%%%%%%%%%%%%%%%%
% DATA AVAILABILITY STATEMENT
%%%%%%%%%%%%%%%%%%%%%%%%%%%%%%%%%%%%%%%%%%%%%%%%%%%%%%%%%%%%%%%%%%%%%
% 
%
\paragraph{Data availability statement}

The data used to generate all of the figures in this work is publicly available at the following link: \url{https://doi.org/10.5281/zenodo.13931577
}. 
%  The data availability statement is where authors should describe how the data underlying 
%  the findings within the article can be accessed and reused. Authors should attempt to 
%  provide unrestricted access to all data and materials underlying reported findings. 
%  If data access is restricted, authors must mention this in the statement. See
%  {http://www.ametsoc.org/PubsDataPolicy} for more info.

%%%%%%%%%%%%%%%%%%%%%%%%%%%%%%%%%%%%%%%%%%%%%%%%%%%%%%%%%%%%%%%%%%%%%
% APPENDIXES
%%%%%%%%%%%%%%%%%%%%%%%%%%%%%%%%%%%%%%%%%%%%%%%%%%%%%%%%%%%%%%%%%%%%%
%
%% If only one appendix, use

%\appendix

%% If more than one appendix, use \appendix[<letter>], e.g.,

%\appendix[A] 

%% Appendix title is necessary! For appendix title:

%\appendixtitle{Title of Appendix}

%%% Appendix section numbering (note, skip \section and begin with \subsection)
%
% \subsection{First primary heading}

% \subsubsection{First secondary heading}

% \paragraph{First tertiary heading}

%%%%%%%%%%%%%%%%%%%%%%%%%%%%%%%%%%%%%%%%%%%%%%%%%%%%%%%%%%%%%%%%%%%%%
% REFERENCES
%%%%%%%%%%%%%%%%%%%%%%%%%%%%%%%%%%%%%%%%%%%%%%%%%%%%%%%%%%%%%%%%%%%%%
% Make your BibTeX bibliography by using these commands:
\printbibliography

\end{document}